\documentclass[pra,a4paper,twocolumn,superscriptaddress,longbibliography,nofootinbib]{revtex4-2}

\usepackage{amssymb}
\usepackage{amsmath}
\usepackage{amsfonts}
\usepackage{graphicx}
\usepackage{bm}
\usepackage{xcolor}
\usepackage{multirow}
\usepackage{comment}
\usepackage{subfigure} 

\usepackage{amsmath}
\usepackage{graphicx}
\usepackage{bm}
\usepackage{xcolor}
\usepackage{comment}

\definecolor{darkblue}{rgb}{0.1,0.2,0.6}
\definecolor{darkred}{rgb}{0.8,0.1,0.2}
\definecolor{darkgreen}{rgb}{0,0.6,0.1}
\usepackage[colorlinks,citecolor=darkblue,linkcolor=darkred,urlcolor=darkblue]
{hyperref}
\usepackage[all]{hypcap}
\usepackage[normalem]{ulem}
\usepackage{mathtools}
\usepackage{amsmath}%
\usepackage{amsfonts}%
\usepackage{amssymb}
\newcommand{\bg}{ \begin{gather} }
\newcommand{\eg}{\end{gather}}
\newcommand{\be}{ \begin{equation} }
\newcommand{\ee}{\end{equation}}
\newcommand{\bea}{ \begin{eqnarray} }
\newcommand{\eea}{\end{eqnarray}}

\begin{document}

\title{Disorder--driven  transition to tubular phase  in anisotropic     two-dimensional materials
}

\author{M. V. Parfenov}

\affiliation{Department of Physics, HSE University, 101000 Moscow, Russia}

\affiliation{Laboratory for Condensed Matter Physics, HSE University, 101000 Moscow, Russia}

\author{V. Yu. Kachorovskii}

\affiliation{Ioffe  Institute, 194021 St.~Petersburg, Russia}

\author{I. S. Burmistrov}

\affiliation{\hbox{L. D. Landau Institute for Theoretical Physics, Semenova 1-a, 142432, Chernogolovka, Russia}}

\affiliation{Laboratory for Condensed Matter Physics, HSE University, 101000 Moscow, Russia}


\begin{abstract}
We develop a  theory of anomalous elasticity in disordered two-dimensional flexible materials with  orthorhombic crystal symmetry. Similar to the clean case, we predict existence of infinitely many flat phases with anisotropic bending rigidity and Young's modulus showing power-law scaling with momentum controlled by a single universal exponent the very same as in the clean isotropic case. With increase of temperature or disorder these flat phases undergo crumpling transition.  
Remarkably, in contrast to the isotropic 
materials where crumpling occurs in all spatial directions simultaneously, the anisotropic materials crumple into tubular phase. In distinction to clean case in which crumpling transition happens  at unphysically high temperatures,  a 
disorder-induced tubular crumpled phase can exist even at room-temperature conditions.  Our results are applied to anisotropic atomic single layers doped by adatoms or disordered by heavy ions bombarding. 
\end{abstract}

\maketitle


\section{Introduction} 

The discovery of graphene~\cite{Novoselov2004,Novoselov2005,Zhang2005} and other atomically  thin  materials \cite{Novoselov2012} opened the  field of flexible two-dimensional (2D)  materials, 
the so-called crystalline membranes \cite{2Dmat}. 
 A hallmark of such membranes is anomalous  elasticity, i.e. non-trivial  scaling  of elastic modules with the system size \cite{Nelson1987}. Also, 2D materials 
 undergo crumpling with increasing  temperature \cite{Nelson1987,Aronovitz1988,Paczuski1988,David1988,Aronovitz1989,Guitter1988,Guitter1989,Doussal1992}. However, the critical temperature  of the crumpling transition (CT) in clean  isotropic membranes  is unphysically high (of order of a few eV). As it has recently been demonstrated \cite{Gornyi:2015a,Saykin2020b}, the CT can occur at low temperature in a disordered  isotropic membrane provided disorder is higher than a certain critical value.  In this paper, we demonstrate that   crystalline anisotropy adds new physics into the problem. We find that with increasing disorder the crumpling occurs in anisotropic way, so that a disordered membrane undergoes transition to the so-called \emph{tubular crumpled} phase. This transition can happen at room or even lower temperature.              
 
There are many examples of anisotropic crystalline membranes, although  
 the hexagonal crystal symmetry of graphene seems to be high enough to make elasticity and electronic transport to be identical to those of isotropic materials. On the other hand, now researchers are interested in many other 2D atomically thin materials, in particular, 2D black phosphorus (phosphorene) \cite{Ling2015,Galluzzi2020}, metal monochalcogenide \cite{Sarkar2020,Barraza-Lopez2021}  and dichalcogenide \cite{Wang2018,Durnev2018} monolayers, etc. These novel 2D materials have low crystal symmetry and, thus, can demonstrate anisotropic physical properties, including elastic response, electron and thermal transport, photolumenescence, Raman scattering, optical absorption, etc.

Probably the most studied one among novel anisotropic atomically thin materials is a transition metal dichalcogenide monolayer that has $D_{3h}$ point symmetry group as opposed to $D_{6h}$ in graphene. However, elastic properties of a transition metal dichalcogenide monolayer is identical to that of graphene, i.e. to an isotropic crystalline membrane~\footnote{For the point symmetry group $D_{3h}$ ($D_{6h}$) the second rank symmetric tensor $u_{jk}$ is transformed according to  the following irreps: $A_1^\prime$ ($A_{1g}$): $u_{xx}+u_{yy}$, $u_{zz}$; $E^\prime$ ($E_{2g}$): $u_{xx}-u_{yy}$, $u_{xy}$; $E^{\prime\prime}$ ($E_{1g}$): $u_{xz}$, $u_{yz}$. Therefore, the elastic energies for $D_{3h}$ and $D_{6h}$ are identical.}. In contrast, phosphorene, boat- and washboard-graphane \cite{Colombo2011}, metal monochalcogenide monolayers (SiS, SiSe, GeS, GeSe, SnS, SnSe), monolayers GeAs$_2$, WTe$_2$, ZrTe$_5$, and Ta$_2$NiS$_5$ have the orthorhombic crystal symmetry \cite{Li2019} such that their elastic response does not reduce to that of an isotropic crystalline membrane.

Let us recap the basic facts on the anomalous elasticity.
It's key ideas were put forward for isotropic crystalline membranes  and dates back to the seminal paper~\cite{Nelson1987}. The developed field theoretical treatment of thermal fluctuations was used to demonstrate the existence of two distinct phases: the low-temperature  flat phase  and the high-temperature isotropic crumpled phase separated by the CT  \cite{Aronovitz1988,Paczuski1988,David1988,Aronovitz1989,Guitter1989,Doussal1992}. Physics behind the CT is the competition between  thermal fluctuations which tend to crumple membrane  and   anharmonicity-induced increase of  bending rigidity with the system size  $L$  that  can  stabilize the membrane  in the flat phase  at $L{\to}\infty.$   Although CT occurs at unphysically high temperatures, the anomalous elasticity
  manifests itself also deep in the  flat phase. In particular,  Young's modulus and bending rigidity in the flat phase have anomalous power-law scaling with $L$   
(or, equivalently, momentum) that leads to nonlinear Hooke's law, negative Poisson ratios, etc. Currently there is a substantial interest in further theoretical understanding of physics of clean isotropic crystalline membranes \cite{Kats2014,Gornyi:2015a,Kats2016,Burmistrov2016,Gornyi2016,Kosmrlj2017,Doussal2018,Burmistrov2018a,Burmistrov2018b,Saykin2020,Saykin2020b,Coquand2020,Mauri2020,Mauri2021,Mauri2022,Metayer2022}.

An extension of the anomalous elasticity theory for  {\it  anisotropic} crystalline membranes has been done in Ref. \cite{Toner1989}. A field theoretic analysis for a membrane of $D{=}4{-}\epsilon$ dimension (with
$\epsilon{\ll}1$) demonstrated that the membrane  becomes asymptotically isotropic at large enough length scales (the so-called universal regime, see below). Thus effective elastic response of such anisotropic membranes should be equivalent to that of an isotropic crystalline membrane. Recently, two of us have shown that this is not the case for an orthorhombic crystalline membrane with the physical  dimension $D{=}2$ \cite{Burmistrov2022}. In the universal regime such an anisotropic membrane has a discrete hidden symmetry preserving the degree of orthorhombicity. 
As a membrane size tends to infinity, $L{\to}\infty$, the discrete symmetry transforms into an emergent continuous symmetry that controls
anisotropy effects in elastic response of an orthorhombic 2D membrane.  With increasing the temperature a 2D membrane with the orthorhombic crystal symmetry undergoes the transition into the tubular crumpled phase --- an anisotropic phase predicted earlier for strongly anisotropic materials \cite{Radzihovsky1995,Radzihovsky1998}.  We notice however that  critical temperature of the transition to the tubular phase is unphysically high for experimentally studied clean 2D anisotropic crystalline membranes.

Realistic 2D flexible materials are disordered  due to  random imperfection of the crystal lattice. The degree of  disorder can be increased by doping with adatoms or by bombarding  of  membrane by heavy ions \cite{Giordanelli2016}.   As a result, 
in addition to thermal fluctuations, the so-called ripples --- the static, frozen deformations --- exist. Similarly to thermal fluctuations disorder-induced ripples affect elastic response and tend to crumple the membrane \cite{Morse:1992,Nelson_1991,Radzihovsky1991,Morse:1992b,Bensimon_1992,Gornyi:2015a,Gornyi2016,Saykin2020b}. An interplay of ripples and thermal fluctuations makes the physics of disordered membranes to be much richer than that of the clean ones. The properties of disordered  membranes are not well understood and are actively discussed both theoretically and experimentally. In particular,  the relevance of disorder for 2D flexible materials has recently been proved by
experimental measurements of nonlinear Hooke's law in graphene \cite{Nicholl2015,Nicholl2017}. These  results are substantially different from the ones predicted for the generic clean membranes  theoretically \cite{Guitter1988,Guitter1989,Aronovitz1989} and  numerically for clean graphene \cite{Los2016}.   
The experimental results can be explained by the one-loop renormalization group (RG) theory of disordered membrane \cite{Gornyi2016}.   
The experimental observations   can be also interpreted as  existence of other flat phase (so-called rippled flat phase) in 2D flexible materials, which reveals itself within two-loop RG analysis \cite{Saykin2020b}. 
Additionally, numerical simulations of graphene clearly demonstrated the disorder-induced CT \cite{Giordanelli2016}. It is worth stressing that the disorder-induced CT in isotropic membranes  happens isotropically, so that for a certain critical value of disorder the membranes simultaneously shrinks in all directions.  \cite{Gornyi2016,Giordanelli2016, Saykin2020b}

Initially, theoretical studies of disordered 2D  membranes predicted the existence of the marginal rippled flat phase at not too high temperatures  within one-loop RG analysis \cite{Morse:1992b,Bensimon_1992}. The scaling of elastic properties  of realistic disordered finite-size  membranes is well described by this marginal  phase even at room temperature up to a very large values of $L.$
However, this phase is unstable and disappears in the thermodynamic limit, $L{\to}\infty$. 
(Similar prediction has been proposed 
for a disordered membrane of dimension $D{=}4{-}\epsilon$ \cite{Morse:1992,Nelson_1991,Radzihovsky1991}). 

Moreover, recently, two of us demonstrated 
that within two-loop RG analysis the marginal rippled phase is stabilized by sufficiently large disorder \cite{Saykin2020b}. This, in turn, means   
the existence of the transition between clean and rippled flat phases at finite temperature (and/or disorder) in 2D disordered membranes. 
Similar conclusion about the existence of the transition between rippled and clean flat phases has been drawn in analysis of a disordered $D{=}4{-}\epsilon$ dimensional membrane
\cite{Coquand2018,Coquand2021,Coquand2022}.

As it was demonstrated by experiments in graphene \cite{Nicholl2015,Nicholl2017}, 
 disorder dramatically changes the elastic response of 2D isotropic flexible materials.  
 Evidently, this implies that disordered  anisotropic membranes  would show rich physics that is  very  different from the physics of clean anisotropic 2D materials.       
In particular,
there are several important physical questions: (i) existence of a marginal flat rippled phase 
within simplest one-loop approximation, 
(ii)  stabilization of this phase at finite temperature within two-loop approximation,   
and (iii) disorder-induced transition to tubular crumpled  phase. We are not aware of any study of these questions in the literature. Here we shall focus on the study  of issues (i) and (iii).

In this paper we develop the theory of anomalous elasticity in disordered 2D flexible materials with orthorhombic crystal symmetry. We focus on the universal regime when the typical size of the membrane is large 
in comparison with the so-called Ginzburg scale.
We perform one-loop RG analysis of  disordered membranes with orthorhombic crystal symmetry. We employ the simplest model of disorder that has the same crystalline symmetry as the bending rigidity. 

For sufficiently weak disorder the amplitude of ripples decreases with increasing the system size and in the thermodynamic limit it becomes negligible as compared to the  temperature-induced out-of-plane deformations. Hence, weak disorder is irrelevant and the large-size membrane becomes in the flat clean phase. Similar to recently discussed clean case \cite{Burmistrov2022}, there are  infinitely many clean anisotropic flat phases. The continuous parameter that distinguishes different flat phases  is related to the degree of orthorhombicity of a membrane. 
These phases have anisotropic bending rigidity and Young's modulus, cf. Eq. \eqref{eq:bending:final:FP:1}, as well as anisotropic spatial behavior of roughness correlation functions, cf. Eq. \eqref{eq:bending:final:FP:2}. 

\begin{figure}
\centering
		  	\includegraphics[width=0.4\textwidth]{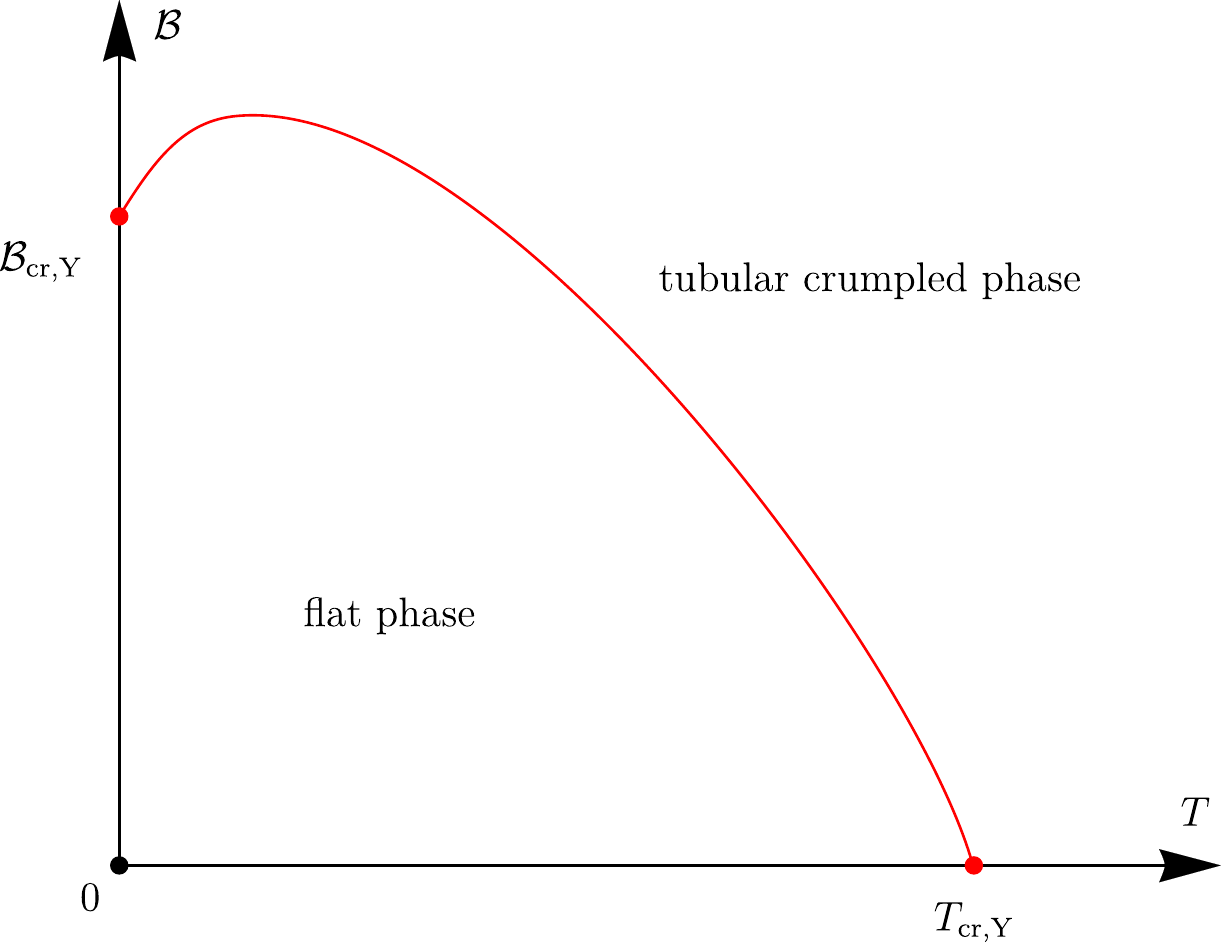}
		  	\caption{Phase diagram of an anisotropic membrane for $\gamma{>}1$. $\mathcal{B}$ is a disorder strength, $T$ is temperature. Solid red curve corresponds to the critical curve, $\mathcal{B}(T)$, separating the flat phase from  
		   the tubular phase of a membrane crumpled in $y$ direction. For $\gamma{<}1$ crumpling occurs in $x$ direction.  
          }\label{pic: phasediagram} 
\end{figure}

We also demonstrate that at large disorder there exists infinite number of  marginal rippled phases. If the disorder strength is smaller than a certain critical value, any marginal phase exists in large but finite interval of scales but in the thermodynamic limit smoothly  transforms to one among  clean  anisotropic flat phases.  However,  if disorder exceeds critical value (different for different  marginal phases),  a marginal phase undergoes transition to a crumpled   phase.   Remarkably,   by contrast to disorder-driven transition  in the isotropic membrane \cite{Gornyi:2015a},  the crumpled phase is \emph{tubular}, so that the  CT is anisotropic and occurs along a certain direction (see Fig.~\ref{pic: phasediagram}).      

What is also dramatically important, especially in view of the experimental application,  is that  in contrast to the clean case  disorder-induced transition to the  tubular crumpled phase can occur at realistic temperatures (at room or even lower temperatures). 

Also we briefly discuss a model of disorder having different symmetry than that of bending rigidity.    In this case we find that the  parameter that distinguishes different flat marginal and tubular crumpled phases may be not directly related with the degree of orthorhombicity of a membrane. 

The outline of is paper is as follows. We formulate the model of 2D anisotropic crystalline membrane in Sec. \ref{Sec:Model}. In Sec. \ref{Sec:Renorm:IM} we perform one-loop renormalization of the free energy in the universal regime (below the inverse Ginzburg length) for the invariant manifold. The derived RG equations are analysed in Sec. \ref{Sec:RGflow:IM}. 
In Sec. \ref{Sec:Crumpling} we study the transition to the tubular phase. The RG flow away from the invariant manifold is studied in Sec. \ref{Sec:RGflow:NIM}. 
We end the paper with discussions and conclusions in Sec. \ref{Sec:Disc}. Some technical details are summarized in Appendices.

\section{Model\label{Sec:Model}} 

We start from the free energy for thermal fluctuations of a  2D membrane in $d{=}3$ dimensional space.  
We assume that the membrane is in the flat phase and has orthorhombic crystal symmetry. Then the free energy acquires the following form \cite{Toner1989}
\begin{gather}
\mathcal{F} = \frac{1}{2}\int_{\bm{x}} \Bigl [\varkappa_{\alpha\beta} \nabla_\alpha^2 \bm{r}\nabla_\beta^2 \bm{r}+ 
c_{\alpha\beta} u_{\alpha\alpha} u_{\beta\beta} 
+ 4c_{66}u_{xy}^2 \Bigr ].
\label{eq:action:0}
\end{gather}
Here, $\int_{\bm{x}}{=}\int d^2 \bm{x}$ and $u_{\alpha\beta} {=} (\partial_\alpha \bm{r}\partial_\beta \bm{r}{-}\delta_{\alpha\beta})/2$ where $\alpha,\beta=1,2$. The point on the membrane is parametrized by a $d{=}3$ dimensional vector $\bm{r}$. 

The four parameters $\{c_{11}, c_{12}, c_{22}, c_{66}\}$ denote the elastic moduli of  a 2D layer of a crystalline material with the orthorhombic crystal symmetry. In the case of $\varkappa_{xx}{=}\varkappa_{yy}$ and $c_{11}{=}c_{22}$, the crystal symmetry is promoted to the tetragonal one. For graphene which has the hexagonal symmetry, the bending energy is isotropic,  $\varkappa_{xx}{=}\varkappa_{yy}{=}\varkappa_{xy}$ together with $c_{11}{=}c_{22}{=} \lambda{+}2\mu$, $c_{12}{=}\lambda$, and $c_{66}{=}\mu$. 

Deformation of the membrane is given as the sum of the  homogeneous and inhomogeneous contributions. In-plane homogeneous
stretching 
is described by the tensor $\xi_{\alpha\beta},$ which 
is proportional to the unit matrix for a clean membrane at zero temperature. 
For a sake of simplicity, we do
not discuss shear deformations here. We thus assume that $\xi_{\alpha\beta}$ has
two nonzero spatially independent diagonal components:
$\xi_{xx}=\xi_x$ and $\xi_{yy}=\xi_y.$ These global deformations  play the key role in the CT. Due to coupling with out-of-plane displacements, they decrease with increasing both temperature (because of  increase of the thermal fluctuations)  and disorder (because of increase of the ripple's amplitude). The CT occurs when one of these  deformations turns to zero. In the isotropic  case,   $\xi_x=\xi_y,$ and CT means shrinking of the global deformation to the point. By contrast, in the anisotropic case,  one of the 
stretchings 
vanishes first, that implies the CT to the tubular phase.

Separating homogeneous stretching, we choose a standard parametrization of the coordinates: $r_1{=}\xi_x x + u_x$, $r_2{=}\xi_y y +u_y$, and $r_{3}{=}h$, such that $u_{\alpha\beta}= (\xi_\alpha^2-1)\delta_{\alpha\beta}/2+\tilde{u}_{\alpha\beta}$, where (no summation over repeating indices is assumed)
\begin{equation}
\tilde{u}_{\alpha\beta}= \frac{1}{2}\Bigl (
\xi_\beta\partial_\alpha u_\beta +\xi_\alpha\partial_\beta u_\alpha+\partial_\alpha h \partial_\beta h + \partial_\alpha \bm{u}\partial_\beta \bm{u} \Bigr ).
\label{eq:u:alpha:beta}
\end{equation}
The 
inhomogeneous deformation  is given by the sum of the in-plane displacement
 $\bm{u}{=}\{u_x,u_y\}$ and the out-of-plane deformation $h.$ Under assumption that the membrane is not too close to the crumpling transition from the flat phase, the term  $\partial_\alpha \bm{u}\partial_\beta \bm{u}$  in Eq. \eqref{eq:u:alpha:beta} can be neglected in comparison with $\partial_\alpha h\partial_\beta h$. Then, the free energy becomes Gaussian with respect to the in-plane displacements. Following Ref. \cite{Nelson1987}, we integrate over $\bm{u}$ and obtain the effective free energy written in terms of the out-of-plane phonons alone,
\begin{align}
\mathcal{F} = & \frac{1}{8}\! \int_{\bm{x}}\, c_{\alpha\beta} \varepsilon_\alpha\varepsilon_\beta +  \frac{1}{2}\int_{\bm{x}} \, \left(\varkappa_{\alpha\beta} \nabla_\alpha^2 h \nabla_\beta^2 h + \beta \nabla^2 h \right)
 \notag \\ + & \frac{1}{8} \int_{\bm{q}} Y(\theta_{\bm{q}}) 
 \Biggl | \int_{\bm{k}} [\bm{k}\times\hat{\bm{q}}]^2 h_{\bm{k+q}}h_{-\bm{k}}\Biggr |^2 .
\label{eq:action:2}
\end{align}
Here we use a short-hand notation, $\int_{\bm{q}}{=}\int \frac{d^2\bm{q}}{(2\pi)^2}$, and introduce $\hat{\bm{q}}{=}\bm{q}/q$ that is the unit vector along the vector $\bm{q}$. Here 
$\varepsilon_\alpha{=}\xi_\alpha^2{-}1{+}  \int_{\bm{k}}
k^2_\alpha h_{\bm{k}}h_{-\bm{k}}$
is displacement, which contains anomalous contribution   $\int_{\bm{k}}
k^2_\alpha h_{\bm{k}}h_{-\bm{k}}$   responsible for   
the anomalous Hooke's law. 
The `prime' sign in the last integral in Eq. \eqref{eq:action:2} indicates that the interaction with $q{=}0$ is excluded. Effective coupling between out-of-plane modes is given by the angle-dependent function 
\begin{equation}
Y(\theta_{\bm q}) =c_{66}\Bigl [\hat{q}_x^2\hat{q}_y^2+\frac{c_{66}
c_{\alpha\beta}\epsilon_{\alpha\alpha^\prime}\epsilon_{\beta\beta^\prime}}{c_{11}c_{22}-c_{12}^2} \hat{q}_{\alpha^\prime}^2 \hat{q}_{\beta^\prime}^2 \Bigr ]^{-1} ,
\end{equation}
where $\epsilon_{\alpha\beta}$ is  fully antisymmetric tensor.  
In the isotropic case, this function does not depend on the angle and is given by the Young modulus  $4 \mu (\lambda{+}\mu)/(\lambda{+}2 \mu).$  Hence,  $Y(\theta_{\bm q})$ represents the   bare value of the  anisotropic Young's modulus \cite{Wei2014}.

The term $(1/2) \int_{\bm{x}}\beta \nabla^2 h $ in Eq. \eqref{eq:action:2} is responsible for a disorder of ``random-curvature'' type. 
 A standard model of such disorder implies that $\beta(\bm {x} )$ has isotropic Gaussian  distribution \begin{gather}
 P_{\rm iso}\{ \beta (\bm{x})\}=\mathcal{N}^{-1} e^{-\int_{\bm{x}} \beta^2(\bm{x})/4\lambda_0}=  \mathcal{N}^{-1} e^{-\int_{\bm{k}}  \beta_{\bm{k}} \beta_{\bm{-k}}/4\lambda_0}.
\label{eq:P:dist:iso} 
\end{gather}
 Here   $\mathcal N$ is the normalization coefficient and  $\lambda_0$ characterizes  the  disorder strength.  For the distribution \eqref{eq:P:dist:iso}, the   correlation function in the  momentum space is isotropic: $ \overline{\beta_{\bm{k}}\beta_{-\bm{k^\prime}}}{=}4 \lambda_0\, \delta_{\bm{k},\bm{k^\prime}} .$    However, as we shall demonstrate below, the  RG flow forces this correlation function to become anisotropic in the momentum space:    
 \begin{equation}
\overline{\beta_{\bm{k}}\beta_{-\bm{k^\prime}}}=4\lambda(\theta_{\bm{k}})\, \delta_{\bm{k},\bm{k^\prime}} ,
\qquad 
\lambda(\theta)=  \hat k_\alpha^2 \lambda_{\alpha\beta} \hat k_\beta^2 .
\label{beta-theta}
\end{equation}
Here  we introduced  symmetric tensor $\lambda_{\alpha \beta}$  instead of  single variable   $\lambda_0$. Such correlation function is consistent with the RG flow. 
In the isotropic case all $\lambda_{\alpha \beta}$ are equal to $\lambda_0$. This implies that the RG flow generates  two additional  coupling constants, cf. Eq. \eqref{eq:psi:def:harm}.
We also notice that Eq.~\eqref{beta-theta} is reproduced by  distribution  function 
\begin{gather}
 P\{ \beta (\bm{k})\}= \mathcal{N}^{-1} e^{-\int_{\bm{k}}  \beta_{\bm{k}} \beta_{\bm{-k}}/4\lambda(\theta_{\bm{k}}) } .
\label{eq:P:dist:non-iso} 
\end{gather}
We stress that the distribution function changes its functional form under the RG.
We assume that the function $\lambda (\theta)$ in Eq.~\eqref{beta-theta} is non-negative for all angles, i.e., \color{black}
\begin{equation}
    \lambda_{xx}>0, \quad \lambda_{yy}>0, \quad \lambda_{xy}>-\sqrt{\lambda_{xx}\lambda_{yy}} .
    \label{eq:cond:lambda}
\end{equation}

It is convenient to introduce  bare angle-dependent bending rigidity 
\begin{equation}
\varkappa(\theta_{\bm{k}})=\varkappa_{\alpha\beta}\hat{k}_\alpha^2\hat{k}_\beta^2=\varkappa_0 +\varkappa_2 \cos(2 \theta_{\bm{k}}) + \varkappa_4 \cos(4 \theta_{\bm{k}}), 
\label{kappa-theta}
\end{equation}
where $\theta_{\bm{k}}$ is the angle of the wave vector $\bm{k}$ and $\varkappa_0 {=} (3 \varkappa_{xx} {+} 2 \varkappa_{xy} {+} 3 \varkappa_{yy})/8,$
$\varkappa_2{=}(\varkappa_{xx}{-}\varkappa_{yy})/2,$ and $\varkappa_4{=} (\varkappa_{xx} {-} 2 \varkappa_{xy} {+} \varkappa_{yy})/8.$  Generally, there are two anisotropic terms characterized by bending rigidities  $\varkappa_2$ and $\varkappa_4$. In the case of the tetragonal crystal symmetry, the second harmonics proportional to $\varkappa_2$ is absent.

In what follows we assume that the following  inequalities hold 
\begin{equation}
\varkappa_{xx}{>}0, \quad \varkappa_{yy}{>}0 , \quad \varkappa_{xy}{>}{-}(\varkappa_{xx} \varkappa_{yy})^{1/2} . 
\label{eq:assumpt:kappa}
\end{equation}
They guarantee  that $\varkappa(\theta_{\bm{k}})$ is positive for all angles $\theta_{\bm{k}}$. Consequently,  the membrane is stable against transition into a tubular phase at zero temperature and in the absence of disorder.  

Next we make two more adjustments of the effective free energy \eqref{eq:action:2}. At first, we introduce $N$ replica in order to be able to perform averaging of $\ln \mathcal{F}$ over disorder. Secondly, we extend the dimensionality of the membrane's embedding space $d$ from $3$ to $2{+}d_c$. Additional dimension $d_c$ plays a role of the flavor index $N_{\rm f}$ for out-of-plane phonons. Below we use standard approach, analogous to $1/N_{\rm f}$ expansion over number of flavors:  we assume that additional dimension is large  and 
use perturbation theory controlled by parameter  $1/d_c{\ll} 1.$ All in all, we substitute the scalar field $h$ by a tensor field $h_{j}^{(a)}$ where $j=1,\dots, d_c$ and $a=1,\dots N$. In what follows we shall use the vector notation $\bm{h}^{(a)}=\{h_{1}^{(a)},\dots,h_{d_c}^{(a)}\}$.
Then, after averaging over disorder the replicated free energy becomes
\begin{align}
\mathcal{F} & =   \frac{1}{8}\! \int\limits_{\bm{x}} \sum_{a=1}^N c_{\alpha\beta}  \varepsilon_\alpha^{(a)}\varepsilon_\beta^{(a)} +  \frac{1}{2}\sum_{a,b=1}^N \Bigl\{
\int\limits_{\bm{k}} \,  \varkappa_{ab}(\theta_{\bm{k}}) k^4  \bm{h}^{(a)}_{\bm{k}} \bm{h}^{(b)}_{-\bm{k}} 
 \notag \\ + & \frac{1}{4}\int\limits_{\bm{q}} Y(\theta_{\bm{q}}) \delta_{ab}
 X_{\bm{q}}^{(a)} X_{-\bm{q}}^{(b)} \Bigr \}, \,\, X_{\bm{q}}^{(a)}{=}\int\limits_{\bm{k}}[\bm{k}{\times}\hat{\bm{q}}]^2 \bm{h}^{(a)}_{\bm{k+q}}\bm{h}^{(a)}_{-\bm{k}}
\label{eq:action:21}
\end{align}
where $\varepsilon_\alpha^{(a)}{=}\xi_\alpha^2{-}1{+} \! \int_{\bm{k}}
k^2_\alpha \bm{h}^{(a)}_{\bm{k}}\bm{h}^{(a)}_{-\bm{k}}$. The  quantities $\varkappa_{ab}(\theta)$ are the elements of the $N\times N$ matrix in the replica space,
\begin{equation} \label{eq: bendrig: def}
\hat{\varkappa}(\theta) = \varkappa(\theta) \hat{1}- \psi(\theta) \hat{J} ,
\end{equation}
where $\hat{J}$ is the identity matrix, $J_{ab}{=}1$. 
The function $\psi(\theta)$ is defined as follows
\begin{equation}
\psi(\theta_{\bm{k}}) = \frac{1}{T}\hat{k}_\alpha^2 \lambda_{\alpha \beta} \hat{k}_\beta^2 .
\end{equation}
Similar to $\varkappa(\theta)$, cf. Eq.~\eqref{kappa-theta}, the function $\psi(\theta_{\bm{k}})$ can be expanded in the Fourier series 
\begin{equation}
\psi(\theta)  =\psi_0+ \psi_2 \cos(2\theta)+\psi_4 \cos(4\theta) .
\label{eq:psi:def:harm}
\end{equation}
Here we introduce $\psi_k{=} \lambda_k/T$  with $k{=}0, 2, 4$ and harmonics
$\lambda_0 {=} (3 \lambda_{xx}{+}2 \lambda_{xy}{+}3 \lambda_{yy})/8,$
$\lambda_2{=}(\lambda_{xx}{-}\lambda_{yy})/2,$ and $\lambda_4{=}(\lambda_{xx}{-}2 \lambda_{xy}{+} \lambda_{yy})/8.$ 
Below we will also use notation $\psi_{\alpha \beta}= \lambda_{\alpha \beta}/T.$

\section{Renormalization on the invariant manifold\label{Sec:Renorm:IM}}

Generically, there are no relations between components of matrices $\varkappa_{\alpha\beta}$ and $\psi_{\alpha\beta}$. Inspired by the emergent symmetry in the clean case \cite{Burmistrov2022}, at first we assume that the following relation holds
\begin{equation}
\left(\frac{\varkappa_{xx}}{\varkappa_{yy}}\right)^{1/4}= \left(\frac{\psi_{xx}}{\psi_{yy}}\right)^{1/4} \equiv \gamma ,
\label{eq:relation:gamma}
\end{equation}
where $0{<}\gamma{<}\infty$. The parameter $\gamma$ controls asymmetry between $x$ and $y$ axes existing  in the orthorhombic symmetry class. We note that in the case of tetragonal crystal symmetry the relation \eqref{eq:relation:gamma} holds trivially with $\gamma{=}1$.

Below we shall demonstrate that the RG flow preserves  Eq.~\eqref{eq:relation:gamma}. A more general case when   Eq.~\eqref{eq:relation:gamma} does not hold for bare values of $\varkappa_{\alpha\beta}$ and $\psi_{\alpha\beta}$ is discussed in Sec.~\ref{Sec:RGflow:NIM}.

\subsection{Elimination of the second harmonics}

We use the same approach as it was used in Ref.~\cite{Burmistrov2022} for analysis of the clean case. Specifically, we  eliminate the second angular harmonic  of angle-dependent bending rigidity $\varkappa(\theta)$ and disorder function $\psi(\theta),$  having in mind to  reduce the problem to analysis of the system with tetragonal symmetry.       

The condition \eqref{eq:relation:gamma} allows one to eliminate the second angular harmonics 
in both $\varkappa(\theta)$ and $\psi(\theta)$, simultaneously. We perform the affine transformation of momenta and coordinates
\begin{equation}
k_x \mapsto  k_x/\sqrt{\gamma}, \, k_y\mapsto k_y \sqrt{\gamma}, \quad 
x\mapsto x \sqrt{\gamma}, \, y \mapsto y/\sqrt{\gamma} . 
\label{eq:kxky:transform}
\end{equation}
Making this transformation we find that the free energy \eqref{eq:action:21}
keeps the same form but the bending rigidities, Young's modulus, and displacements become modified. The functions $\varkappa(\theta)$ and  $\psi(\theta)$ transform as
\begin{equation}
\begin{split}
\varkappa(\theta) \mapsto \widetilde{\varkappa}(\theta) & =
\widetilde{\varkappa}_0+\widetilde{\varkappa}_4 \cos(4\theta) , \\
\psi(\theta) \mapsto \widetilde{\psi}(\theta) & =
\widetilde{\psi}_0+\widetilde{\psi}_4 \cos(4\theta) .
\end{split}
\label{eq:kappa:new}
\end{equation}
Here the zeroth and fourth angular harmonics are expressed in terms of components of the bending rigidity and disorder function as follows
\begin{gather}
\widetilde{\varkappa}_0 = \frac{\sqrt{\varkappa_{xx}\varkappa_{yy}}}{1+t},\, 
\widetilde{\varkappa}_4 = t \widetilde{\varkappa}_0, 
 \,
\widetilde{\psi}_0 = \frac{\sqrt{\psi_{xx}\psi_{yy}}}{1+s}, \, \widetilde{\psi}_4 = s \widetilde{\psi}_0
\notag \\
t = \frac{\sqrt{\varkappa_{xx}\varkappa_{yy}}-\varkappa_{xy}}{3\sqrt{\varkappa_{xx}\varkappa_{yy}} +\varkappa_{xy}}, \,  s = \frac{\sqrt{\psi_{xx}\psi_{yy}}-\psi_{xy}}{3\sqrt{\psi_{xx}\psi_{yy}} +\psi_{xy}}.
\label{eq:kappa:new:2}
\end{gather}
It is worth noting that in the case of the tetragonal crystal symmetry, when $\varkappa_{xx}{=}\varkappa_{yy} {\neq} \varkappa_{xy}$  and 
 $\psi_{xx}{=}\psi_{yy} {\neq} \psi_{xy},$ we have $\tilde \varkappa_0{=}\varkappa_0$ and $\tilde \psi_0{=}\psi_0$, although $t{\neq} 0$ and $s{\neq} 0$. 
We also note that the assumptions \eqref{eq:assumpt:kappa} restrict the parameter $t$ to be within the range $|t|{<}1$. It describes the tetragonal distortion of the bending energy of the membrane.  Similarly, conditions \eqref{eq:cond:lambda} restrict the values of $s$ to the interval $|s| {<} 1$.

After the affine transformation \eqref{eq:kxky:transform} is performed, the Young's modulus becomes
\begin{gather}
Y(\theta) \mapsto \widetilde{Y}(\theta)= 4 c_{66}\Bigl [\sin^2(2\theta)+\frac{4c_{66}}{c_{11}c_{22}-c_{12}^2} \Bigl (\frac{c_{11}}{\gamma^2} \cos^4\theta
 \notag \\
 - \frac{c_{12}}{2}\sin^2(2\theta) +c_{22} \gamma^2\sin^4\theta\Bigr )
 \Bigr ]^{-1} .
\end{gather}
Similarly, after affine transformation \eqref{eq:kxky:transform} the displacements become
\begin{equation}
\begin{split}
\varepsilon_x^{(a)} \mapsto \widetilde{\varepsilon}_x^{(a)}
{=} &
\xi_x^2 
{-}1{+}\frac{1}{\gamma}
\int_{\bm{k}}
 k_x^2
\bm{h}^{(a)}_{\bm{k}}\bm{h}^{(a)}_{-\bm{k}} , \\
\varepsilon_y^{(a)} \mapsto \widetilde{\varepsilon}_y^{(a)}
{=} &
\xi_y^2 
{-}1{+}\gamma
\int_{\bm{k}}
 k_y^2
\bm{h}^{(a)}_{\bm{k}}\bm{h}^{(a)}_{-\bm{k}} .
\end{split}
\label{deformations}
\end{equation}

We stress that although the orthorhombicity parameter $\gamma$ disappears from the bending part of the free energy, the Young's modulus  as well as deformations $\widetilde{\varepsilon}_{x,y}^{(a)}$ depend explicitly on it. Thus the symmetry of the free energy $\mathcal{F}$ after the rescaling \eqref{eq:kxky:transform}  is still lower than the tetragonal one.  In the next section we shall demonstrate that screening of the interaction between  out-of-plane modes removes the constant $\gamma$ from the effective coupling in the universal regime. Hence,  the  symmetry of the free energy increases up to tetragonal one. In turn, it implies that there exists a hidden symmetry in the problem in a full analogy with the clean case \cite{Burmistrov2022}. Hence, there are infinite number of phases characterized by parameter $\gamma$ which is not changed under RG flow
\cite{Burmistrov2022}:
\be
\gamma={\rm const}\, .
\ee
Remarkably, the parameter $\gamma$ is still involved in Eqs.~\eqref{deformations}, so that the orthorhombical  crystal symmetry is of crucial importance for the CT to the tubular phase.

\subsection{Screening of interaction}
\begin{figure}[b]
\centering \subfigure{\includegraphics[scale=0.42]{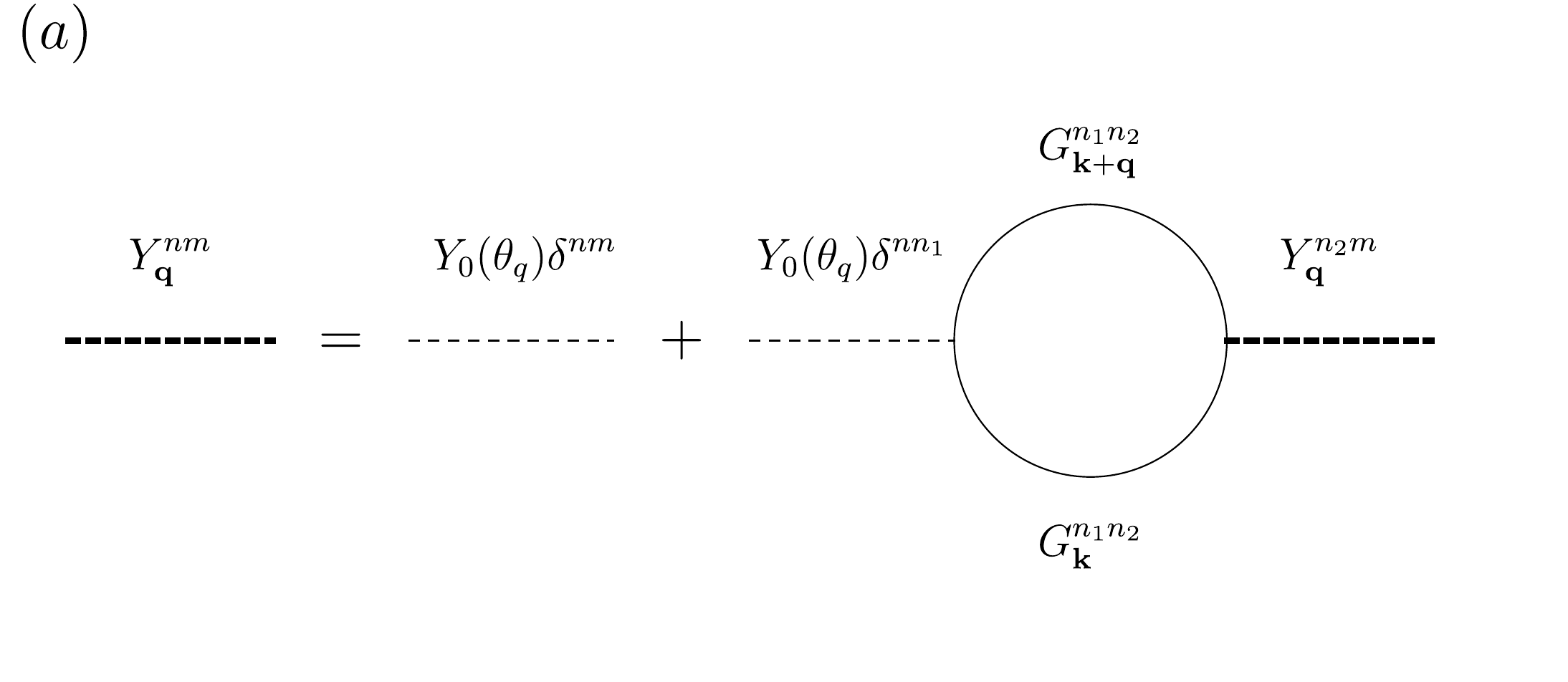}}
	\qquad
	\centering \subfigure{\includegraphics[scale=0.45]{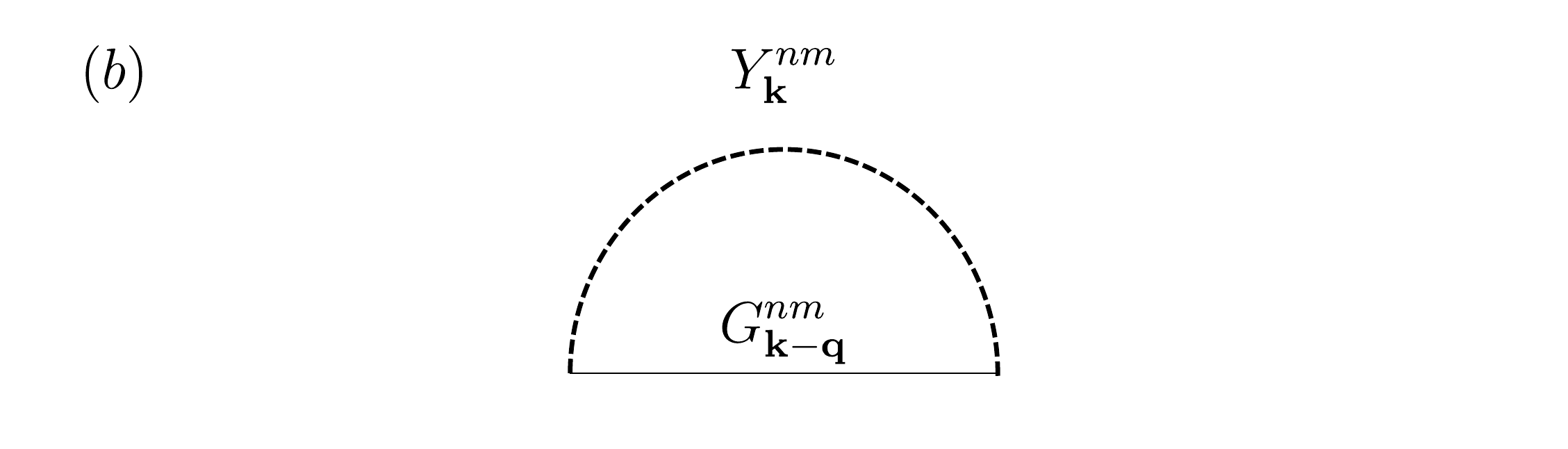}}
		  	\caption{(a) The RPA-type resummation for the interaction.
(b) The one-loop self-energy correction. The
solid line represents the bare Green's function. The thin
(thick) dashed line denotes the (bare) screened interaction. The superscripts denote indices in the replica space.}\label{pic: diagram} 
\end{figure}

In this and next subsections we are working in the rescaled frame of reference. For a sake of brevity we shall not use `tilde' sign below for quantities in that rescaled coordinate system. Information on renormalization of the effective bending rigidity $\varkappa(\theta)$ and disorder function $\psi(\theta)$ can be extracted from the exact two-point Green's function  
\begin{equation}
\langle h_i^{(a)}(\bm{k}) h_j^{(b)}(-\bm{k})\rangle {\equiv} \mathcal{G}_{ab}(\bm{k}) \delta_{ij} ,
\end{equation}
where the average is taken with respect to the free energy $\mathcal{F}$. The quadratic part of $\mathcal{F}$ determines the bare Green's function 
\begin{equation}
\hat{G}(\bm{k}) {=} \frac{T}{\varkappa(\theta_{\bm k}) k^4} \Bigl (\hat{1}+f(\theta_{\bm k}) \hat{J}\Bigr ), \qquad f(\theta)=\frac{\psi(\theta)}{\varkappa(\theta)} .
\end{equation} 

As usual, the bare interaction, $Y(\theta)$, between flexural phonons is screened by diagrams of RPA-type shown in Fig.~\ref{pic: diagram}. In particular, after taking such screening into account the diagonal in replica space interaction, cf. Eq. \eqref{eq:action:21},
 becomes non-diagonal one,
\begin{equation}
Y(\theta_{\bm{q}})\hat{1} \mapsto \hat{Y}(\theta_{\bm{q}})  = Y(\theta_{\bm{q}}) \Bigl (\hat{1}+3 Y(\theta_{\bm{q}})\hat{\Pi}(\bm{q})/2\Bigr )^{-1} .
\label{eq:Y-RPA}
\end{equation}
In the leading order in small parameter $1/d_c$ the polarization operator is given by a bare bubble:
\begin{gather}
\Pi_{ab}(\bm{q}) {=} \frac{d_c}{3T}\int\limits_{\bm{k}} [\bm{k}{\times}\hat{\bm{q}}]^4
G_{ab}(\bm{k})G_{ab}(\bm{k+q}) {=} \frac{d_c T}{\varkappa_0^2 q^2}\mathcal{P}_{ab}(\theta_{\bm{q}}) ,
\label{eq:Pi-Bubble}
\end{gather}
where the matrix $\hat{\mathcal{P}}(\theta_{\bm{q}})$ is dimensionless. As one can see from Eqs. \eqref{eq:Y-RPA} and \eqref{eq:Pi-Bubble}, the
screened interaction becomes independent of $Y(\theta_{\bm{q}})$ in the long wave limit,  $q{\ll}q_*$ \cite{Nelson1987,Aronovitz1988,Aronovitz1989}.  Here the inverse Ginzburg length can be estimated $q_*{\sim} 
(\sqrt{d_c Y}/\varkappa)\max\{T, \lambda\}$, where $\varkappa$, $Y$, and $\lambda$ are typical values of bending rigidity, Young's modulus, and disorder variance, respectively. 
Additionally, at $q{\ll}q_*$ the interaction between flexural phonons becomes small (${\sim} 1/d_c$) being determined by the inverse polarization operator ${\sim} \hat{\Pi}^{-1}$. 
Consequently, the free energy $\mathcal{F}$  becomes independent of the 
orthorhombicity parameter $\gamma$ as a consequence of emergent  
hidden symmetry. We note that although $\gamma$ remains in the expressions for the displacements ${\varepsilon}_a$, cf. Eqs. \eqref{deformations}, it does not affect renormalization of  ${\varkappa}_{0,4}$ and ${\psi}_{0,4}$.

Before going to the computation of the renormalization of the free energy $\mathcal{F}$, we discuss the polarization operator in more details. It is convenient to represent it as follows 
\begin{align}
\hat{\mathcal{P}}(\theta_{\bm{q}}) {=} & 
\frac{\varkappa_0^2 q^2}{3} \int\limits_{\bm{k}} [\bm{k}{\times}\hat{\bm{q}}]^4
\frac{[1+2f(\theta_{\bm{k+q}})]\hat{1}+f(\theta_{\bm{k+q}})f(\theta_{\bm{k}})\hat{J}}{|\bm{k+q}|^4 k^4 \varkappa(\theta_{\bm{k+q}})\varkappa(\theta_{\bm{k}})}\notag \\
{=} & [\pi_{00}(\theta_{\bm{q}}) + 2 f_\textsf{o} \pi_{10}(\theta_{\bm{q}})]\hat{1}+
f_\textsf{o}^2 \pi_{11}(\theta_{\bm{q}}) \hat{J} .
\end{align}
Here we introduce $f_\textsf{o} = \psi_0/\varkappa_0$ (we emphasize that $f_\textsf{o}$  is not the zeroth angular harmonics of $f(\theta)$).     
 The dimensionless function  $\pi_{nm}(\theta)$ is given as 
\begin{widetext}
\begin{gather}
\pi_{nm}(\theta) {=}\! \int\limits_0^\infty\! \frac{d z z}{6\pi}\! \int\limits_0^{2\pi}\!\frac{d\varphi}{2\pi} \frac{\sin^4(\varphi{-}\theta) 
[\kappa(s,\varphi)]^{n}}{[\kappa(t,\varphi)]^{1{+}n}}
\frac{[(1{-}s)(z^2{+}2z\cos(\varphi{-}\theta){+}1)^2{+}2s(z^2\cos(2\varphi){+}2z\cos(\varphi{+}\theta){+}\cos(2\theta))^2]^m}{[(1{-}t)(z^2{+}2z\cos(\varphi{-}\theta){+}1)^2{+}2t(z^2\cos(2\varphi){+}2z\cos(\varphi{+}\theta){+}\cos(2\theta))^2]^{1{+}m}} ,
\label{eq: pifunc}
\end{gather}
\end{widetext} 
where we introduce $\kappa(t,\varphi){=}1{+}t \cos(4\varphi)$. The function $\pi_{nm}$ 
depends on  $s$ and $t$  and does not depend on $f_\textsf{o}$  [for a sake of brevity we skip  arguments $s$ and $t$ in the notation  $\pi_{nm}(\theta)$].  
We mention that $\pi_{nm}(\theta{-}\pi/2){=}\pi_{nm}(\theta)$. This implies that $\pi_{nm}(\theta)$ can be expanded in the Fourier series in quartic angular harmonics $\cos(4 m \theta)$.
We note that the function $\pi_{10}(\theta)$ can be expressed in terms of $\pi_{00}(\theta)$: 
\begin{equation}
\pi_{10}(\theta)=\pi_{00}(\theta) - (s-t)\partial_t \pi_{00}(\theta)/2 .
\label{pi10}
\end{equation}
We present the discussion of the  behavior of the functions $\pi_{nm}(\theta)$ in various regimes in Appendices \ref{app: ast0} and \ref{app: ast1}.

We emphasize that the variable $f_\textsf{o}$ is proportional to disorder strength and inversely proportional  to temperature. Physically, it describes competition between ripples and thermal fluctuations: for $f_\textsf{o}{\ll} 1$ the contribution of the ripples is negligible  and a membrane can be considered as a clean one.  The opposite limit, $f_\textsf{o}{\gg} 1,$ corresponds to ``dirty'' membrane such that elasticity is fully determined by ripples. As we shall demonstrate below, in this limit temperature drops out from both  the RG equations and the conditions for the CT. In fact, this case  of strong disorder  is formally equivalent to the $T{=}0$ limit  (since  thermal fluctuations   give negligible contribution), although the physical temperature can be sufficiently large.

\subsection{Self-energy}

The smallness of screened interaction allows one to construct the regular perturbation theory in $1/d_c$ for the self-energy $\hat{\Sigma}(\bm k) {=} \hat{G}^{-1}(\bm k){-}\hat{\mathcal{G}}^{-1}(\bm k)$. To the lowest order in $1/d_c$ the self-energy is given as (see diagram in Fig.~\ref{pic: diagram}),
\begin{align}
\Sigma_{ab}(\bm{k})&=  - \int\limits_{\bm{q}} [\bm{k}{\times}\hat{\bm{q}}]^4 Y_{ab}(\bm{q}) G_{ab}(\bm{k-q}) 
\simeq - \frac{2\varkappa_0^2}{3 d_c T} \notag \\
\times & \int\limits_{\bm{q}} \frac{[\bm{k}{\times}\bm{q}]^4}{q^2}
\frac{\delta_{ab}+f(\theta_{\bm{k-q}})J_{ab}}{|\bm{k-q}|^4 \varkappa(\theta_{\bm{k-q}})}
\Bigl (\hat{\mathcal{P}}^{-1}(\theta_{\bm{q}})\Bigr )_{ab} .
\label{eq:SelfEnergy:12}
\end{align} 
In order to extract renormalization of harmonics $\varkappa_{0,4}$ and $\psi_{0,4}$, we need to compute the self-energy in the limit $k{\to}0$. The integral over absolute value of $\bm{q}$ in Eq. \eqref{eq:SelfEnergy:12} is logarithmically divergent with $k$ providing the low energy cut off. With logarithmic accuracy we find at $k{\to} 0$, 
\begin{gather}
\Sigma_{ab}(\bm{k})  \simeq -\frac{\varkappa_0^2}{24\pi d_c T}k^4 \ln \frac{q_*}{k} 
\int\limits_0^{2\pi} \frac{d\theta_{\bm{q}}}{2\pi} \Bigl (3 - 4 \cos 2\theta_{\bm{k}} \cos 2 \theta_{\bm{q}} \notag \\
+ \cos 4 \theta_{\bm{k}} \cos 4 \theta_{\bm{q}}\Bigr ) \frac{\delta_{ab}+f(\theta_{\bm{q}})J_{ab}}{\varkappa(\theta_{\bm{q}})}\Bigl (\hat{\mathcal{P}}^{-1}(\theta_{\bm{q}})\Bigr )_{ab} .
\label{eq:Sigma:ab:1}
\end{gather}
Here we use approximation $\theta_{\bm{k-q}}{\simeq} 
\pi{-}\theta_{\bm{q}}$ for $k{\to} 0$.
Using the inverse of the polarization operator
\begin{gather}
\hat{\mathcal{P}}^{-1} = \frac{\hat{1}}{\pi_{00}+2 f_\textsf{o}\pi_{10}} -
 \frac{f_\textsf{o}^2 \pi_{11} \hat J}{(\pi_{00}+2 f_\textsf{o}\pi_{10})^2} ,
\end{gather}
we obtain
\begin{gather}
\hat{\Sigma}(\bm{k})  \simeq -\frac{\varkappa_0}{24\pi d_c T}k^4 \ln \frac{q_*}{k} 
\int\limits_0^{2\pi} \frac{d\varphi}{2\pi} \Bigl (3 - 4 \cos 2\theta_{\bm{k}} \cos 2 \varphi \notag \\
+  \cos 4 \theta_{\bm{k}} \cos 4 \varphi\Bigr )
\frac{1}{\kappa(t,\varphi)} \frac{1}{\pi_{00}(\varphi) +2 f_\textsf{o} \pi_{10}(\varphi)} 
\notag \\
\times
 \Biggl \{ \Bigl [ 1 + f_\textsf{o} \frac{\kappa(s,\varphi)}{\kappa(t,\varphi)}
 - \frac{f_\textsf{o}^2 \pi_{11}(\varphi)}{\pi_{00}(\varphi) +2 f_\textsf{o} \pi_{10}(\varphi)}\Bigr ] \hat{1}
 \notag \\
 - \frac{f_\textsf{o}^3 \pi_{11}(\varphi)}{\pi_{00}(\varphi) +2 f_\textsf{o} \pi_{10}(\varphi)} \frac{\kappa(s,\varphi)}{\kappa(t,\varphi)} \hat{J}
 \Biggr \} .
 \label{eq:Sigma:Pert}
\end{gather}
The properties of the functions $\kappa(t,\varphi)$ and $\pi_{nm}(\varphi)$ guarantees that the second harmonic of the self-energy, the term in Eq. \eqref{eq:Sigma:Pert} proportional to $\cos 2\theta_{\bm{k}}$, vanishes identically. 

\subsection{Renormalization group equations}

In a standard way, the perturbative correction \eqref{eq:Sigma:Pert} to the self-energy can be translated into the RG equations for harmonics ($m{=}0,2$)
\begin{equation}
\begin{split}
\frac{d\varkappa_{2m}}{d\Lambda}=\frac{2 \varkappa_0 }{(1+m)d_c} F_{2m}(f_\textsf{o},t,s),\\
\frac{d\psi_{2m}}{d\Lambda}=\frac{2\psi_0 }{(1+m)d_c} \Phi_{2m}(f_\textsf{o},t,s) .
\end{split}
\label{eq:RG:1}
\end{equation}
Here $\Lambda=\ln(q_*/k)$ and we introduced the following functions:
\begin{align}
F_{2m}(f_\textsf{o},t,s) & = \frac{1}{16\pi} \int\limits_0^{2\pi} \frac{d\varphi}{2\pi} 
\frac{\cos(2m\varphi)}{\kappa(t,\varphi)} \Biggl [ \frac{1 + f_\textsf{o} \kappa(s,\varphi)/\kappa(t,\varphi)}{\pi_{00}(\varphi) +2 f_\textsf{o} \pi_{10}(\varphi)}
\notag \\
 & - \frac{f_\textsf{o}^2 \pi_{11}(\varphi)}{[\pi_{00}(\varphi) +2 f_\textsf{o} \pi_{10}(\varphi)]^2}\Biggr ]
 \label{eq:F2m:def}
\end{align}
and
\begin{align}
\Phi_{2m}(f_\textsf{o},t,s) & = \frac{1}{16\pi} \int\limits_0^{2\pi} \frac{d\varphi}{2\pi} 
\frac{\cos(2m\varphi)}{\kappa(t,\varphi)} \frac{\kappa(s,\varphi)}{\kappa(t,\varphi)}
\notag \\
& \times \frac{f_\textsf{o}^2 \pi_{11}(\varphi)}{[\pi_{00}(\varphi) +2 f_\textsf{o} \pi_{10}(\varphi)]^2} .
 \label{eq:Phi2m:def}
\end{align}
We note that the functions $F_0$ and $
\Phi_0$ are even under simultaneous change of signs of $t$ and $s$ whereas $F_4$ and $\Phi_4$ are odd,
\begin{equation}
    \begin{split}
F_{2m}(f_\textsf{o},-t,-s) & = (-1)^{m/2} F_{2m}(f_\textsf{o},t,s),   \\ 
\Phi_{2m}(f_\textsf{o},-t,-s) & = (-1)^{m/2} \Phi_{2m}(f_\textsf{o},t,s) .
\end{split}
\label{eq:symm:F:Phi}
\end{equation}

We mention that Eqs.  \eqref{eq:RG:1} are valid to the lowest order in $1/d_c$.
Using Eqs. \eqref{eq:RG:1}, we can write down closed set of RG equations for dimensionless variables, $f_\textsf{o}$, $t$, and $s$:
\begin{align}
\frac{d f_\textsf{o}}{d\Lambda} & =  \frac{2 f_\textsf{o}}{d_c} \Bigl [ \Phi_0(f_\textsf{o},t,s) - F_0(f_\textsf{o},t,s)\Bigr ] ,\notag\\
\frac{d t}{d\Lambda} & =  \frac{2}{d_c} \Bigl [ \frac{1}{3}F_4(f_\textsf{o},t,s) - t F_0(f_\textsf{o},t,s)\Bigr ] \label{eq:RG:2} ,\\
\frac{d s}{d\Lambda} & =  \frac{2}{d_c} \Bigl [ \frac{1}{3}\Phi_4(f_\textsf{o},t,s) - s \Phi_0(f_\textsf{o},t,s)\Bigr ] .\notag 
\end{align} 
As consequence of relations \eqref{eq:symm:F:Phi}, the RG flow is symmetric with respect to simultaneous change of signs of $t$ and $s$.  
One can easily check that these equations  generate nonzero magnitude of $s$ even if one starts from $s{=0}$. Therefore, the RG flow makes the disorder anisotropic [see Eq.~\eqref{beta-theta}] even if initially disorder is isotropic at the ultra-violet scale.     
Below we shall analyse the three-parameter RG flow governed by Eqs. \eqref{eq:RG:2}.

We note that the difference $\Phi_0(f_\textsf{o},t,s){-}F_0(f_\textsf{o},t,s)$ in Eq. \eqref{eq:RG:2} can have both positive and negative sign such that the RG flow of $f_\textsf{o}$ is non-monotonous, contrary to the isotropic case. Below we shall discuss the RG flow in more detail.

\section{RG flow within the invariant manifold\label{Sec:RGflow:IM}}

\subsection{Isotropic clean fixed point}\label{subsec: Isotrp}

The system \eqref{eq:RG:2} has an infrared stable fixed point: $f_\textsf{o}{=}0$, $t{=}0$, and $s{=}{\rm const}$. Although formally it is a line of fixed points since $s$ is an arbitrary constant,  all points on the line are equivalent since $f_\textsf{o}{=}0$.  Physically, this means that for large system size,  the  ripples  have a fixed degree of anisotropy $s$  and their amplitude decreases much faster than the amplitude of the thermal fluctuations. Hence, disorder is irrelevant in the thermodynamic limit, $L{\to}\infty.$ There are infinitely many flat phases  characterized by parameter $\gamma.$ These phases are realized provided that  bare value of disorder is sufficiently small: $f_\textsf{o} {\lesssim}1.$ 

\begin{figure}
\centering
		  	\includegraphics[width=0.4\textwidth]{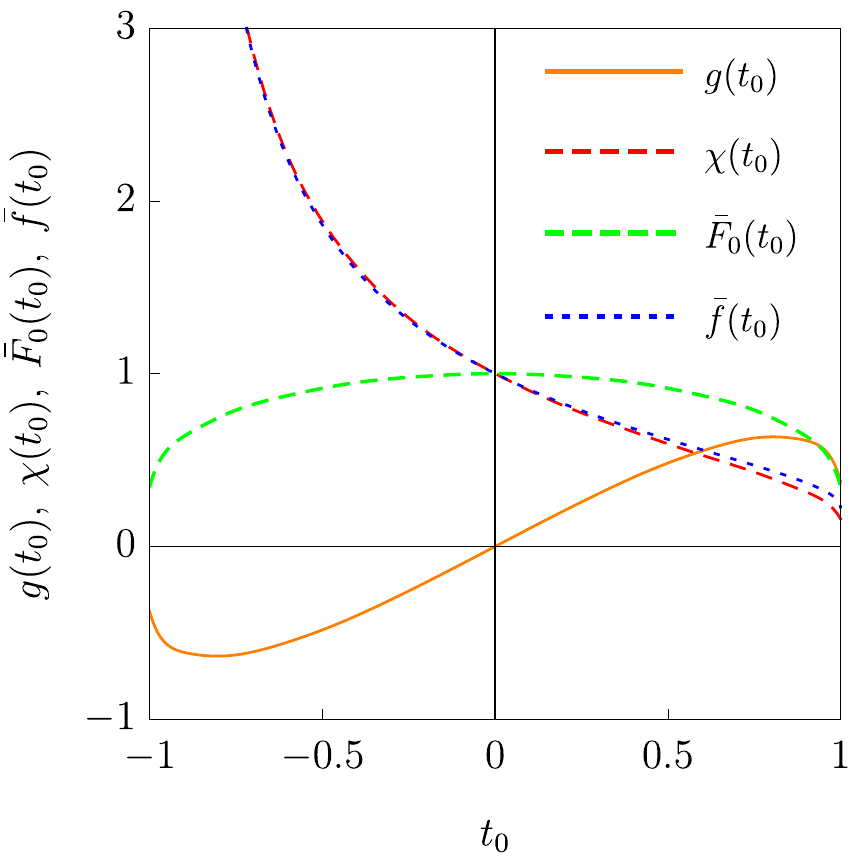}
		  	\caption{Functions $g(t_0),\; \chi(t_0), \; \Bar{F}_0(t_0)$, that determines Eqs. \eqref{eq:RG:fLarge:ts}, \eqref{eq:stretch:st:1} and $\Bar{f}(t_0)$ that determines critical disorder (see Eq. \eqref{eq:stretch:st:3}) }\label{pic: func2} 
\end{figure}

Expanding RG Eqs. \eqref{eq:RG:1} and \eqref{eq:RG:2} at $f_\textsf{o}{\to}0$, we find:
\begin{equation}
\begin{split}
    \frac{df_\textsf{o}}{d\Lambda} & = - \frac{2 f_\textsf{o}}{d_c} \overline{F}_0(t),\qquad 
    \frac{d t}{d\Lambda}  = - \frac{2}{d_c} g(t), \\
    \frac{d s}{d\Lambda} & = 0 + O(f_\textsf{o}^2), \qquad\qquad \frac{d \varkappa_0}{d\Lambda}  = \frac{2}{d_c}\varkappa_0 \overline{F}_0(t) .
\end{split}
\label{eq:RG:L1}
\end{equation}
Here $\overline{F}_{2m}(t)$ is the function $F_{2m}$ in the absence of disorder, 
\begin{equation}
\overline{F}_{2m}(t) =F_{2m}(f_\textsf{o}=0).
\end{equation}
We note that the later is independent of $s$. Also we introduced the function
\begin{equation}
g(t) = t \overline{F}_{0}(t) - \overline{F}_{4}(t)/3 . 
    \label{eq:g:def}
\end{equation}
The behavior of the functions $g(t)$ and $\overline{F}_{0}(t)$ are shown in Fig.~\ref{pic: func2}.
At $|t|{\ll}1$ the functions $\overline{F}_{0}(t)$ and $g(t)$  have the following expansion \cite{Burmistrov2022}
\begin{equation}
    \overline{F}_{0}(t)\simeq 1 - \frac{25}{81}t^2, \quad 
    g(t) \simeq \frac{65 t}{54}\left(1 - \frac{9527 }{35100}t^2\right) .
    \label{eq:asympt:g:F}
\end{equation}
Hence, in a closed vicinity of the fixed point $f_\textsf{o}{=}0$, $t{=}0$, and $s{=}{\rm const}$, we find
\begin{equation}
    \frac{df_\textsf{o}}{d\Lambda} = - \frac{2 f_\textsf{o}}{d_c},\quad  
    \frac{d t}{d\Lambda}  = - \frac{65 t}{27 d_c}, \quad
    \frac{d s}{d\Lambda} = 0, \quad \frac{d \varkappa_0}{d\Lambda} = \frac{2}{d_c}\varkappa_0 .
\end{equation}
As one can see, $t$ and $f_\textsf{o}$ approach zero with decrease of momentum as a power law: $t {\sim} (k/q_{*})^{65/27d_c}$ and $f_\textsf{o} {\sim} (k/q_{*})^{2/d_c}$. We note that these exponents coincide with the ones found in Refs. \cite{Burmistrov2022} and \cite{Gornyi:2015a}, respectively. The zeroth harmonics of the bending rigidity grows as in the isotropic clean case: $\varkappa_0{\sim} (k/q_{*})^{-2/d_c}$ \cite{Nelson1987}. 

\begin{figure*}[t]
\centering
		  	\includegraphics[width=0.356\textwidth]{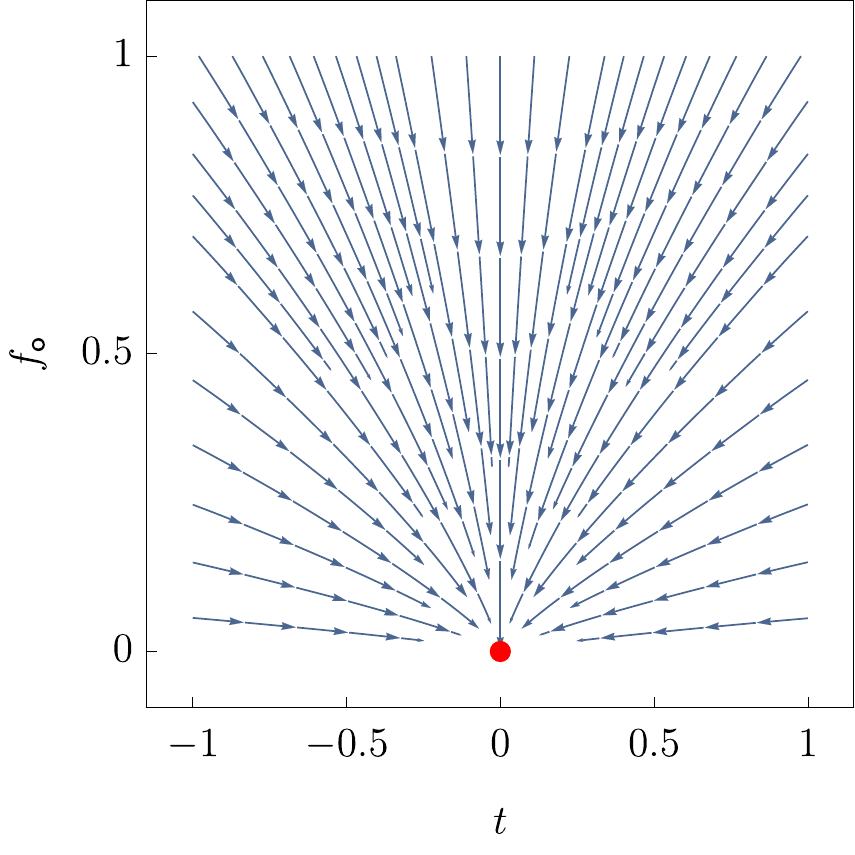}
		  \hspace{0.1\textwidth} \includegraphics[width=0.372\textwidth]{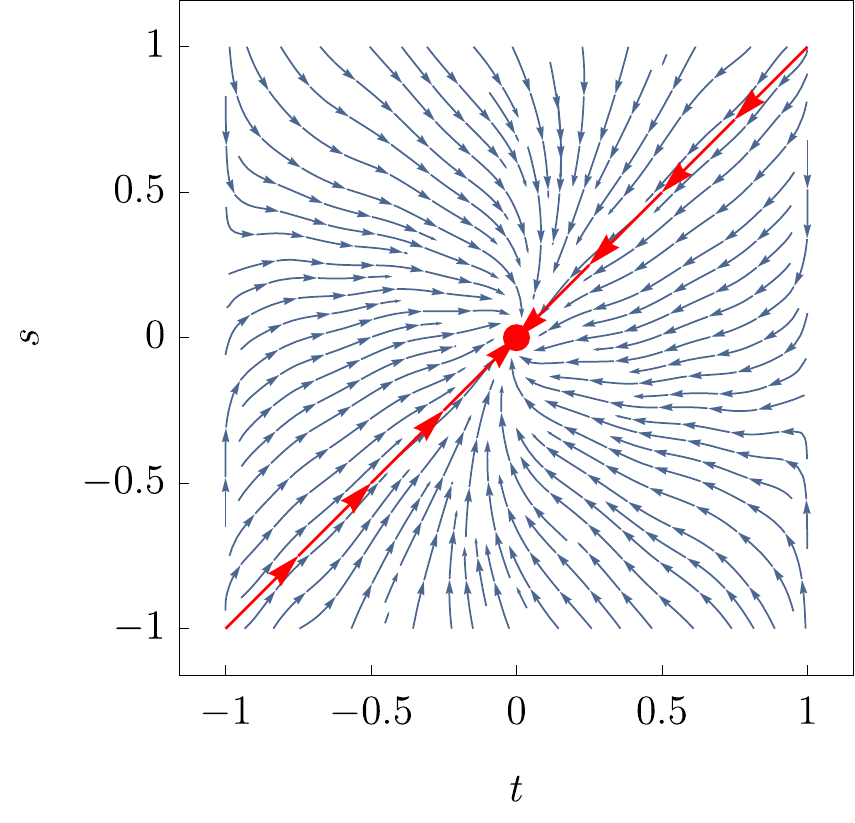}
		  	\caption{RG flow near isotropic clean fixed point $f_\textsf{o}{=}t{=}0$ (left) and  at $f_\textsf{o}{\gg}1$ (right). Solid red line is the line $s{=}t$.}\label{pic: RGflow: 1} 
\end{figure*}

The RG flow around the isotropic clean fixed point is shown in Fig. \ref{pic: RGflow: 1}  (left panel).

\subsection{Renormalization group flow at $f_\textsf{o}{\gg} 1$}

The RG flow in the case of the strong disorder $f_\textsf{o}{\gg} 1$ can be read from Eqs. \eqref{eq:RG:2}, in which the functions $F_{2m}$ and $\Phi_{2m}$ are replaced by their asymptotics $\tilde{F}_{2m}$ and $\tilde{\Phi}_{2m}$, which are independent of $f_\textsf{o}$,
\begin{equation}
\tilde{F}_{2m}(t,s) = \frac{1}{32\pi} \int\limits_0^{2\pi} \frac{d\varphi}{2\pi} 
\frac{\cos(2m\varphi)}{\kappa(t,\varphi)\pi_{10}(\varphi)} \Biggl [ \frac{\kappa(s,\varphi)}{\kappa(t,\varphi)}
 - \frac{\pi_{11}(\varphi)}{2\pi_{10}(\varphi)}\Biggr ]
 \label{eq:F2mf:def}
\end{equation}
and
\begin{equation}
\tilde{\Phi}_{2m}(t,s)  = \frac{1}{64\pi} \int\limits_0^{2\pi} \frac{d\varphi}{2\pi} 
\frac{\cos(2m\varphi)}{\kappa(t,\varphi)} \frac{\kappa(s,\varphi)}{\kappa(t,\varphi)}\frac{\pi_{11}(\varphi)}{\pi^2_{10}(\varphi)} .
 \label{eq:Phi2mf:def}
\end{equation}
Therefore, the RG equations for $t$ and $s$ form a closed system of equations: 
\begin{equation}
\begin{split}
\frac{d t}{d\Lambda} & =  \frac{2}{d_c} \Bigl [ \frac{1}{3}\tilde F_4(t,s) - t \tilde F_0(t,s)\Bigr ] ,\\
\frac{d s}{d\Lambda} & =  \frac{2}{d_c} \Bigl [ \frac{1}{3}\tilde \Phi_4(t,s) - s \tilde\Phi_0(t,s)\Bigr ] , 
\end{split} 
\label{eq:RG:ts}
\end{equation}
while $f_{\textsf{o}}$ obeys
\begin{equation}
\frac{d f_\textsf{o}}{d\Lambda}  =  \frac{2 f_\textsf{o}}{d_c} \Bigl [ \tilde\Phi_0(t,s) - \tilde F_0(t,s)\Bigr ] .
\label{f0bigf0}
\end{equation}
From Eqs.~\eqref{eq:F2mf:def} and \eqref{eq:Phi2mf:def}, one can find 
\begin{equation}
\tilde\Phi_0(t,s) - \tilde F_0(t,s)
= (t-s) R(s,t),  
\label{help1}
\end{equation}
where $R(s,t){\simeq} 5(2s{+}3t)/162$ for $|s|, |t| {\ll} 1$ (see Appendix \ref{app: ast0}). Hence, as follows from Eqs.~\eqref{f0bigf0} and \eqref{help1}, the
variable $f_\textsf{o}$ has very slow change for $|s-t|{\ll} 1$. Thus we can consider it as a constant. Physically, it means existence of  infinite number of marginal phases  distinguished  by the continuous  parameter $\gamma.$  
These phases exist in a wide interval of scales where  $f_\textsf{o}$ can be considered as constant. 
However,  if a bare value of disorder is smaller than a certain critical value (different for different phases), in the thermodynamic limit we finally get $f_\textsf{o}{\to}0$  and arrive at one of the anisotropic flat phases.  If a bare magnitude of disorder exceeds critical value,  the system undergoes transition into tubular crumpled phase as will be discussed in the next section. 
We note also that for $|s{-}t|{\sim}1$ there is no small parameter that controls existence of the marginal phase.  However, numerical analysis shows that $f_{\textsf{o}}$ changes slower than $t$ and $s$  also generically.

Let us now discuss properties of a marginal phase assuming that $f_{\rm o}{=} {\rm const}.$
The RG flow at $f_\textsf{o}{\gg} 1$ is shown in Fig. \ref{pic: RGflow: 1}  (right panel). It has an interesting character. 
Using definition of $\pi_{nm}$, see Eq. \eqref{eq: pifunc}, one can immediately check that at $s{=}t$ all three polarization operators are identical: $\pi_{00}(t,t,\varphi){=}\pi_{10}(t,t,\varphi){=}\pi_{11}(t,t,\varphi)$. Consequently, at $s{=}t$ the functions $\tilde{F}_{2m}$ and $\tilde{\Phi}_{2m}$ coincide: $\tilde{\Phi}_{2m}(t,t){=}\tilde{F}_{2m}(t,t){=}\overline{F}_{2m}(t)/4$. Therefore, it follows that the RG equations for $s$ and $t$ become identical, i.e.  
$s{=}t$ is the invariant line of the RG flow at $f_\textsf{o}{\gg} 1$.\footnote{We note that the condition $s{=}t$ implies the following relation between coefficients of matrices $\psi_{\alpha\beta}$ and $\varkappa_{\alpha\beta}$,
$\psi_{xy}/\sqrt{\psi_{xx}\psi_{yy}}{=}\varkappa_{xy}/\sqrt{\varkappa_{xx}\varkappa_{yy}}$.}

The behavior of $t$ and $\varkappa_0$ on the line $s{=}t$ are governed by the following RG equations
\begin{equation}
\begin{split}
\frac{dt}{d\Lambda}= & -\frac{1}{2d_c} g(t), \qquad \frac{d\varkappa_0}{d\Lambda} = \frac{1}{2d_c} \overline{F}_0(t) \varkappa_0 .
\end{split}     
\label{eq:RG:fLarge:ts}
\end{equation}
We emphasize that Eqs. \eqref{eq:RG:fLarge:ts} transforms into RG equations for the clean case of Ref. \cite{Burmistrov2022} with the replacement $d_c{\to} d_c/4$. 

As one can see from Eqs. \eqref{eq:RG:fLarge:ts} the variable $t$ flows towards zero, i.e. the line $s{=}t$ is not the line of fixed points. Moreover, the line $s{=}t$ is not always attractive for the RG flow. Indeed, to the lowest order in the difference $s{-}t$ one finds  
\begin{equation}
\frac{d(s-t)}{d\Lambda} = -\frac{1}{2d_c}g^\prime(t) (s-t) .
\end{equation}
The derivative $g^\prime(t)$ is positive only for $|t|{\lesssim} 0.8$. Hence, the line $s{=}t$ is attractive line for the RG flow around the fixed point $t{=}s{=}0$.  
Since at $|t|{\ll}1$ the function $g(t)$ is linear, the exponent that controls approaching the line $s{=}t$ coincides with the exponent at which $t$ approaches zero along the line, $s{-}t{\sim} (k/q_{*})^{65/108d_c}$ and $t {\sim}  (k/q_{*})^{65/108d_c}$. These features are clearly seen in the RG flow shown in Fig. \ref{pic: RGflow: 1} (right panel).

As one can see from the right panel on Fig. \ref{pic: RGflow: 1}, an interesting pattern of the RG flow occurs for $1{-}|t|{\ll} 1$. Since RG flow is symmetric with respect to inversion $t{\to}{-}t$ and $s{\to}-s$ we concentrate on the region $1{-}t{\ll}1$. At first, we remind that at $t{=}1$ the bending rigidity vanishes along the lines in the momentum space $k_x{=}\pm k_y$ ($\theta_{\bm k}{=}\pm \pi/4$). Therefore, one can expect singularities at $t{=}1$ (and $s{\neq} t$) in functions $\tilde{F}_{2m}$ and $\tilde{\Phi}_{2m}$, see Fig. \ref{fig: F, Phi}. For $1{-}t{\ll}1$ and $1{-}t{\ll}1{-}s$, we find (see Appendix \ref{app: ast1})
\begin{equation}
\tilde{F}_0 =- \tilde{F}_4\simeq c_{F} , \qquad \tilde{\Phi}_0 =- \tilde{\Phi}_4 \simeq c_{\Phi} \frac{1-s}{1-t} ,
\label{eq:F:Phi:0:0}
\end{equation}
where
\begin{equation}
\begin{split}
c_{F}  & = \frac{3}{2\pi} \int\limits_0^{\pi/2}d\phi\, \frac{3-4\sin^2\phi}{1+\phi \tan\phi} \approx 0.75 , \\
c_{\Phi} & =  \frac{3}{2\pi} \int\limits_0^{\pi/2}d\phi\, \frac{\cos^2\phi}{1+\phi \tan\phi} \approx 0.29 .
\end{split}
\end{equation}
Using asymptotic expressions \eqref{eq:F:Phi:0:0}, we rewrite RG Eqs. \eqref{eq:RG:2} for $1{-}t{\ll}1$ and $1{-}t{\ll}1{-}s$ as
\begin{gather}
  \frac{d f_\textsf{o} }{d\Lambda} =  \frac{2 c_{\Phi}}{d_c} f_\textsf{o}\frac{1-s}{1-t}, \quad   \frac{d t}{d\Lambda}  = -\frac{8 c_{F}}{3 d_c} , \quad \frac{d\varkappa_0}{d\Lambda}=\frac{2 c_F}{d_c}\varkappa_0 , \notag \\
    \frac{d s}{d\Lambda} = -\frac{2 c_{\Phi}}{3d_c} \frac{(1-s)(1+3s)}{1-t} . 
     \label{eq: RGtsim1}
\end{gather}
This system can be solved analytically (see Appendix \ref{app: solt1}). We note that $t$ flows away from its initial value $t_0$ logarithmically, $t{=}t_0{-}[8 c_{F}/(3d_c)] \ln q_*/k$. There exits the stable line of fixed points at $s{=}{-}1/3$ which is clearly seen in Fig. \ref{pic: RGflow: 1} (right panel). We note that this `line' is limited to the close vicinity of $t{=}1$. Away from $s{=}{-}1/3$ the parameter $s$ flows slower than $t$, $s{\sim} (\ln q_*/k)^\alpha$ where $\alpha{=}c_{F}/c_{\Phi}{\approx} 0.39$. The magnitude of the parameter $f_\textsf{o}$ changes with the same velocity, $f_\textsf{o}{\sim} (\ln q_*/k)^\alpha$.

\begin{figure}
	\centering \subfigure[]{\includegraphics[scale=0.45]{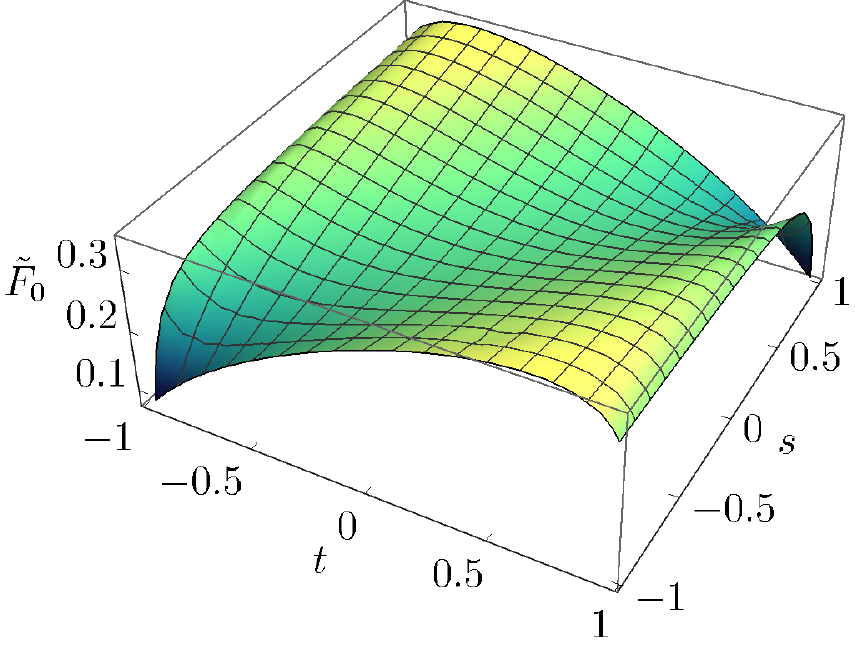}}\qquad 
	\subfigure[]{ \includegraphics[scale=0.45]{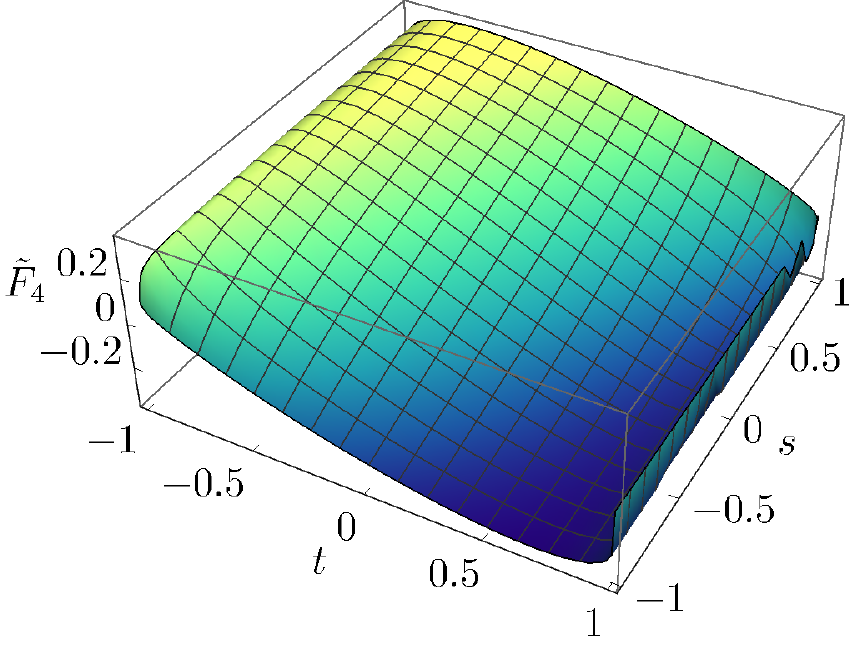}} 
	\qquad
	\centering \subfigure[]{\includegraphics[scale=0.45]{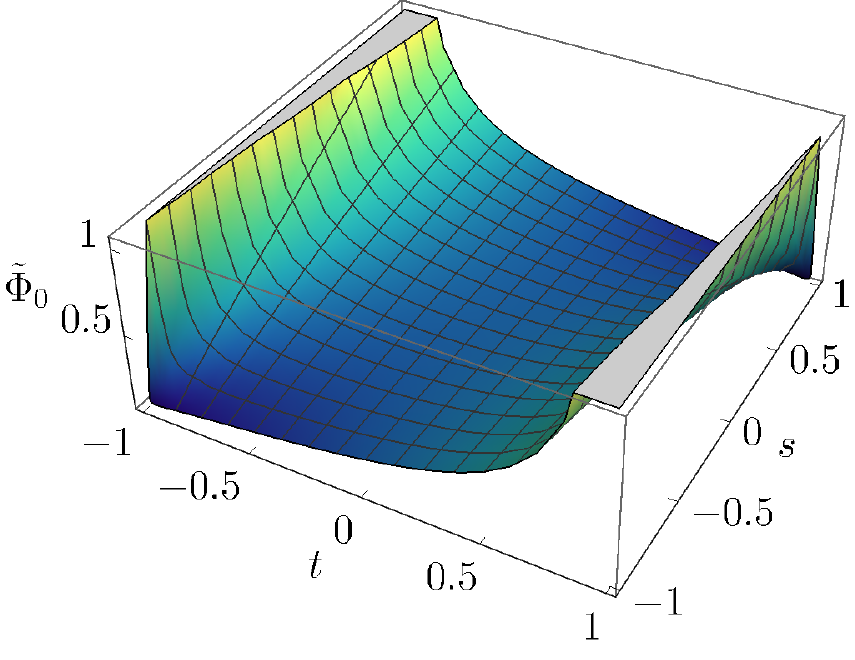}}\qquad 
	\subfigure[]{ \includegraphics[scale=0.44]{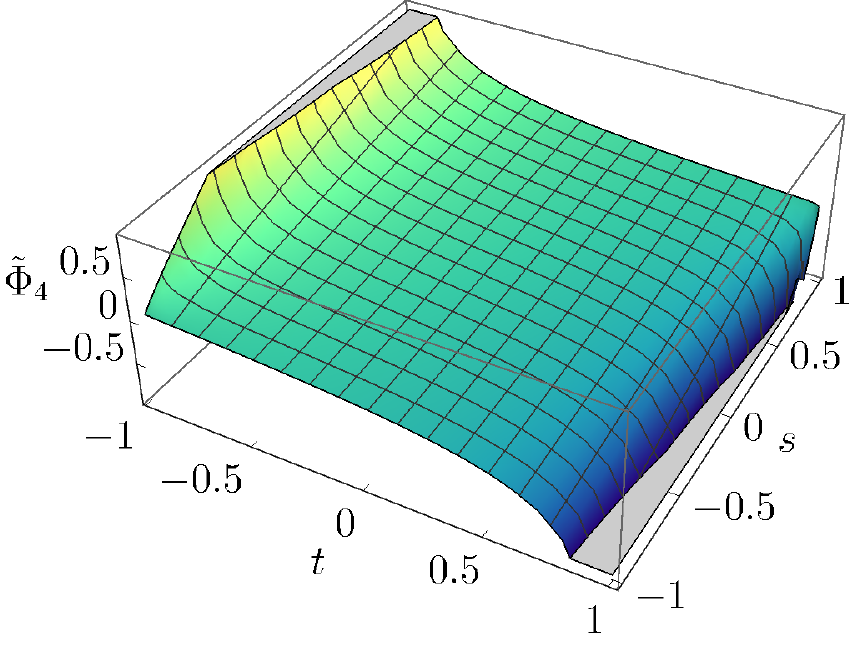} }
	\caption{Dependence of the functions $\tilde{F}_{2m}$ and $\tilde{\Phi}_{2m}$ (cf. Eqs. \eqref{eq:F2mf:def} and \eqref{eq:Phi2mf:def}) on $t$ and $s$ at $f_\textsf{o}{\gg} 1$: (a)  $\tilde{F}_{0}$, (b)  $\tilde{F}_{4}$, (c) $\tilde{\Phi}_{0}$, and (d) $\tilde{\Phi}_{4}$.}
	\label{fig: F, Phi}
\end{figure}

\section{Transition to the tubular crumpled phase in a ``dirty'' membrane 
\label{Sec:Crumpling}}

In the absence of external tension the average displacements are zero, $\langle \varepsilon_\alpha^{(a)} \rangle {=} 0$. These conditions yield equations for the stretching factors in the original (not rescaled) frame of reference,
\begin{equation}
\xi^2_x = 1 - \frac{d_c}{\gamma} \int\limits_{\bm{p}} p_x^2 \mathcal{G}_{aa}(\bm{p}) , \quad
\xi^2_y = 1 - {d_c}{\gamma} \int\limits_{\bm{p}} p_y^2 \mathcal{G}_{aa}(\bm{p}) .
\label{eq:xi:def:1}
\end{equation}
We note that there is no summation over replica indices. 

Following Ref. \cite{Gornyi:2015a}, it is convenient to introduce momentum dependent stretching factors which are given by Eq. \eqref{eq:xi:def:1} with momentum $p$ restricted to be larger than some infrared momentum scale $k$. 
Neglecting renormalization of the Green's function we find logarithmically divergent contributions due to fluctuations of out-of-plane phonons,
\begin{gather}
\begin{pmatrix}
\xi^2_x\\
\xi_y^2
\end{pmatrix} = 1 
- \frac{d_c T}{2\pi\varkappa_0} \ln \left (\frac{q_*}{k}\right ) 
\int \limits_{0}^{2\pi}\frac{d\theta}{2\pi} 
\frac{1}{\kappa(t,\theta)} \Bigl [ 1+ f_\textsf{o} \frac{\kappa(\psi,\theta)}{\kappa(t,\theta)}\Bigr ]
\notag \\ 
\times
\begin{pmatrix}
\gamma^{-1}\cos^2\theta \\
\gamma \sin^2\theta
\end{pmatrix} .
\label{eq:xiXY:full}
\end{gather}
Performing integral over the angle $\theta$ we cast the above equations in the form of the RG equations
\begin{equation}
\begin{split}
\frac{d\xi_x^2}{d\Lambda} & = - \frac{d_c T}{4\pi \gamma \varkappa_0\sqrt{1-t^2}}
\Bigl [1 +f_\textsf{o} \frac{1-s t}{1-t^2}\Bigr ] , \\
\frac{d\xi_y^2}{d\Lambda} & = - \frac{d_c T\gamma }{4\pi \varkappa_0\sqrt{1-t^2}}
\Bigl [1 +f_
\textsf{o} \frac{1-s t}{1-t^2}\Bigr ]  . 
\end{split}
\label{eq:RG:stretching}
\end{equation}
We supplement these equations by the initial conditions $\xi_x^2(0){=}\xi_y^2(0){=}1$.
Equations \eqref{eq:RG:stretching} together with RG Eqs. \eqref{eq:RG:2} allow us to determine the parameters at which the CT occurs. We say that the membrane is in the flat phase if both $\xi_x^2(\Lambda{\to} \infty)$ and $\xi_y^2(\Lambda{\to}\infty)$ are positive. If only one of the stretching factors, e.g. $\xi_x^2$ vanishes at some finite RG time $\Lambda_c$, $\xi_x^2(\Lambda_c){=}0$, the membrane is in the tubular crumpled phase. The membrane is in the crumpled phase if both $\xi_x^2$ and $\xi_y^2$ become zero simultaneously, $\xi_x^2(\Lambda_c){=}\xi_y^2(\Lambda_c){=}0$. 
The later occur in the case $\gamma{=}1$ alone.


General solution of Eqs. \eqref{eq:RG:2} and \eqref{eq:RG:stretching} is complicated, therefore, we shall discuss some interesting limiting cases below.

In the absence of disorder, $f_\textsf{o}{=}0$, Eqs. \eqref{eq:RG:2} and \eqref{eq:RG:stretching} were derived and studied in Ref. \cite{Burmistrov2022}. It was found that the CT to the tubular phase is possible but occurs at unphysically high temperatures of the order of typical values  of the bending rigidity (${\sim}$1 eV). Weak disorder $f_\textsf{o}{\ll}1$ yields corrections to the CT temperature (see Appendix \ref{weak-disorder}).


Now let us discuss the most interesting case of strong disorder, $f_\textsf{o}{\gg} 1.$ In this case,  Eq.~\eqref {eq:RG:stretching} becomes
\begin{equation}
\begin{split}
\frac{d\xi_x^2}{d\Lambda} & = - \frac{d_c }{4\pi \gamma } {\cal B}
 \frac{(1-s t) \sqrt{1+t}}{(1-t)^{3/2}(1+s)} , \\
\frac{d\xi_y^2}{d\Lambda} & = - \frac{d_c \gamma}{4\pi  } {\cal B}
 \frac{(1-s t) \sqrt{1+t}}{(1-t)^{3/2}(1+s)} . 
\end{split}
\label{eq:RG:stretching-strong}
\end{equation}
Here we introduce
the  dimensionless effective disorder strength  
\be
\mathcal{B} {=} \frac{T \sqrt{\psi_{xx} \psi_{yy}}}{\varkappa_{xx} \varkappa_{yy}}= \frac{ T \psi_0}{\varkappa_0^2} \frac{1+s}{(1+t)^2}=\frac{ \lambda_0}{\varkappa_0^2} \frac{1+s}{(1+t)^2}.
\ee
From Eqs.~\eqref{eq:RG:1} and \eqref{eq:RG:ts} 
we find RG equation for $\mathcal B$:
\be
\frac{d\ln \mathcal B}{d \Lambda} =-\frac{2}{d_{\rm c}}\Omega(t,s),
\label{B-Omega}
\ee
where  
\begin{equation}
\Omega= \frac{2 (\tilde F_4/3 -t  \tilde F_0)}{1+t} + \frac{s \tilde\Phi_0-\tilde\Phi_4/3}{1+s} + 2 \tilde F_0   -\tilde \Phi_0. 
\end{equation}
Equations  \eqref{eq:RG:ts},  \eqref{eq:RG:stretching-strong}, and  \eqref{B-Omega}  represent full set of equations describing CT in the strongly disordered  membrane.    

Several comments are in order here.  
Variable $\mathcal B$ does not depend on  $T,$ so that temperature drops out from  Eqs.~\eqref{eq:RG:ts}, \eqref{eq:RG:stretching-strong}, and \eqref{B-Omega}.   Hence,  crumpling is fully determined by disorder, while the thermal fluctuations can be neglected.  
Since disorder can be easily increased, for example,  by bombarding a 2D material by heavy ions,  then, in contrast to clean materials,  the CT  in a ``dirty'' membrane can be observed for realistic temperatures, e.g. at room temperature.     

Below we consider several limiting cases in which function $\Omega(s,t)$ can be significantly simplified.

\subsection{Small $s$ and $t$:  $t{\ll}1,$ $s{\ll}1.$ }
This is the simplest case,  which allows to capture the underlying physics of the disorder-dominated  CT determined by the following equations
\begin{equation}
\begin{split}
\frac{d\xi_x^2}{d\Lambda} & = - \frac{d_c }{4\pi \gamma } {\cal B}, \\
\frac{d\xi_y^2}{d\Lambda} & = - \frac{d_c \gamma}{4\pi  } {\cal B}, 
\\
\frac{d\ln \mathcal B}{d \Lambda} & =-\frac{1}{2 d_{\rm c}}.
\end{split}
\label{eq:RG:stretching-strong-st-small}
\end{equation}
Here, we took into account that  $\Omega(0,0){=}1/4.$

For isotropic case, $\gamma{=}1,$  these equations  yields  critical value of disorder leading to the CT,  $\mathcal B_{\rm cr}{=}2\pi/d_{\rm c}^2$  obtained  previously in Ref.~\cite{Gornyi:2015a}. For $\gamma{=}1$ the CT occurs simultaneously along both directions: at $\mathcal B{=}\mathcal B_{\rm cr},$ we find $\xi_x{=}\xi_y{=}0$ for $\Lambda{=}\infty.$  
Remarkably, this is not the case for anisotropic membrane.  For example, for $\gamma{>}1,$ the CT happens first along $y-$direction. Specifically,  for 
\be
\mathcal B=\mathcal B_{\rm{ cr,Y}}=  \frac{2\pi}{ \gamma d_{\rm c}^2},
\label{eq:Bcr:st:small}
\ee
the membrane shrinks in $y$-direction, $\xi_y {\to} 0$ for $\Lambda{\to}\infty,$ while $\xi_x$ remains positive.   Hence the membrane undergoes transition into crumpled tubular phase. This  is the central result of our work.

\subsection{The case $s{=}t$}

As we discussed above, the line $s{=}t$ is the invariant line of the RG flow at $f_\textsf{o}{\gg}1$. At $s{=}t$ 
 Eqs.\eqref{eq:RG:stretching} become
\begin{equation}
 \frac{d \xi_x^2}{d\Lambda} = -\frac{d_c \mathcal{B}}{4 \pi \gamma}\sqrt{\frac{1+t}{1-t}},\qquad \frac{d \xi_y^2}{d\Lambda} = -\frac{d_c \mathcal{B} \gamma }{4 \pi  }\sqrt{\frac{1+t}{1-t}}  .
 \label{eq:stretch:st:0}
\end{equation}
The flow of $t$ and $\mathcal{B}$ are governed by the following RG equations
\begin{equation}
\frac{dt}{d \Lambda}  = -\frac{1}{2d_c}g(t), \qquad
    \frac{d\ln \mathcal{B}}{d \Lambda}  = - \frac{1}{2d_c} \chi(t) ,
    \label{eq:stretch:st:1}
\end{equation}
where $\chi(t){=}[\overline{F}_4(t)/3 {+} \overline{F}_0(t)]/(1{+}t)$. We mention that after replacing   $d_c$ with $d_c/4$ and  $\mathcal{B} $ with $ 4T/\sqrt{\varkappa_{xx}\varkappa_{yy}},$ Eqs. \eqref{eq:stretch:st:0}  and  \eqref{eq:stretch:st:1} transform into equations describing the CT in the absence of disorder. Solving Eqs. \eqref{eq:stretch:st:0} and \eqref{eq:stretch:st:1}, we find the critical disorder as (we assume $\gamma{>}1$)
\begin{equation}
\mathcal{B}_{\rm cr,Y} =  \frac{2 \pi}{\gamma d_c^2}\overline{f}(t_0) ,   
\label{eq:stretch:st:3}
\end{equation}
where 
\begin{equation}
\overline{f}(t_0) = \Biggl [
\int\limits_0^{t_0} \frac{dt}{g(t)}\left (\frac{1+t}{1-t}\right )^{1/2} e^{-\int\limits_\tau^{t_0} du \frac{\chi(u)}{g(u)}}
\Biggr]^{-1} .
\end{equation}

The critical disorder $\mathcal{B}_{\rm cr,X}$ for $\gamma{<}1$ can be obtained from Eq. \eqref{eq:stretch:st:3} by substitution of $\gamma$ by $\gamma^{-1}$. The critical disorder at $s_0{=}t_0$ is equal to a quarter of the critical temperature (in units of bending rigidity) in the clean case for the same $t_0$. The same one quarter is known to appear in the isotropic case \cite{Gornyi:2015a}. 
In Appendix \ref{T-correction}
we discuss temperature-induced corrections to the critical disorder.

\section{Renormalization group flow away from the invariant manifold\label{Sec:RGflow:NIM}}

Now we discuss RG flow when the condition \eqref{eq:relation:gamma} is not fulfilled. In this case we cannot nullified second harmonics in $\varkappa(\theta)$ and $\psi(\theta)$ simultaneously.  

In this case we formulate the RG procedure as follows. We start from the free energy defined at the RG scale $\Lambda$. Then we make affine transformation \eqref{eq:kxky:transform} with $\gamma{=}(\varkappa_{xx}/\varkappa_{yy})^{1/4}$. There is no second harmonics of $\varkappa(\theta)$ in the transformed frame of reference. The harmonics of disorder in that coordinate frame are given as
\begin{align}
\tilde{\psi}_0 & = \frac{1}{4}\left (3\frac{\gamma^4+\Gamma^4}{2\gamma^2\Gamma^2} \sqrt{\psi_{xx}\psi_{yy}}
+ \psi_{xy}
\right ) ,\notag \\
\tilde{\psi}_2 & = \frac{\Gamma^4-\gamma^4}{2\gamma^2\Gamma^2}\sqrt{\psi_{xx}\psi_{yy}} , \label{eq:tilde:psi0:1} \\
\tilde{\psi}_4 & = \frac{1}{4}\left (\frac{\gamma^4+\Gamma^4}{2\gamma^2\Gamma^2} \sqrt{\psi_{xx}\psi_{yy}}
- \psi_{xy}
\right ) ,
\notag 
\end{align}
where $\Gamma{=}(\psi_{xx}/\psi_{yy})^{1/4}$. We emphasize the appearance of the second harmonic $\tilde{\psi}_2$. Next we perform the renormalization of the free energy from the RG scale $\Lambda$ to the RG scale $\Lambda-d\Lambda$. The structure of obtained RG equations for $\tilde{\varkappa}_{0,4}$ and $\tilde{\psi}_{0,4}$ coincide with that of Eqs. \eqref{eq:RG:1}. However the functions $F_{2m}$ and $\Phi_{2m}$ depend now on the additional parameter 
$\omega{=}\tilde{\psi}_2/\tilde{\psi}_0$ via a substitution $\kappa(s,\phi) {\to} 1{+}\omega \cos(2\phi){+}s\cos(4\phi)$. Similar account for the second harmonic of $\tilde{\psi}(\theta)$ should be performed in definition of polarization operators $\pi_{nm}(\theta)$, cf. Eq. \eqref{eq: pifunc}. Also the parameter $s{=}\tilde{\psi}_4/\tilde{\psi}_0$ depends explicitly on the ratio $\Gamma/\gamma$. The second harmonic of $\tilde\psi(\theta)$ acquires the RG correction (cf. Eq. \eqref{eq:Sigma:ab:1})
$d\tilde{\psi}_2{=}{-}[8 \tilde{\psi}_0/(3 d_c)] \Phi_2(f_0,t,s,\omega) d\Lambda$.
Due to nonzero $\omega$ the second harmonic of the bending rigidity is also induced, 
$d\tilde{\varkappa}_2{=}{-}[8 \tilde{\varkappa}_0/(3 d_c)] F_2(f_0,t,s,\omega) d\Lambda$.

In order to return the form of the free energy at the RG scale $\Lambda-d\Lambda$ back to its form at the RG scale $\Lambda$ we perform infinitesimal affine transformation \eqref{eq:kxky:transform} with parameter $1{+}d\gamma$ instead of $\gamma$, where $d\gamma{=}d\tilde{\varkappa}_2/[2\tilde{\varkappa}_0 (1{+}t)]$. After this transformation the second harmonic in the bending rigidity nullifies. The difference in zeroth and fourth harmonics of $\varkappa$ between and after the transformation is of the second order in $d\Lambda$ and, thus, can be neglected. After the transformation, both the  zeroth and fourth harmonics of $\tilde{\psi}$ acquire linear in $d\Lambda$ correction, ${-}3 \tilde{\psi}_2 d\gamma/2$ and ${-} \tilde{\psi}_2 d\gamma/2$, respectively. The second harmonic is corrected by the term ${-}2(\tilde{\psi}_0{+}\tilde{\psi}_4)d\gamma$. Taking these corrections into account we obtain the following RG equations ($m{=}0,2$): 
\begin{align}
\frac{d\tilde{\psi}_{2m}}{d\Lambda} &=\frac{2 \tilde{\psi}_0}{(1+m)d_c}\Bigl [ 
\Phi_{2m}(f_\textsf{o},t,s,\omega) +\frac{\omega}{1+t}F_2(f_\textsf{o},t,s,\omega)
\Bigr ] , \notag \\
\frac{d\tilde{\psi}_2}{d\Lambda} &=-\frac{8}{3d_c}\tilde{\psi}_0 \Bigl [ 
\Phi_2(f_\textsf{o},t,s,\omega) -\frac{1+s}{1+t}F_2(f_\textsf{o},t,s,\omega)
\Bigr ] , \notag \\
\frac{d\tilde{\varkappa}_{2m}}{d\Lambda} &=\frac{2 \tilde{\varkappa}_0}{(1+m)d_c} F_{2m}(f_\textsf{o},t,s,\omega) .
\label{eq:RG:omega:full}
\end{align}
We note that the functions $F_{0,4}$ and $\Phi_{0,4}$ are even in $\omega$ whereas the functions $F_2$ and $\Phi_2$ are odd.
From the above equations we derive the following RG equation for the parameter $\omega$:
\begin{align}
\frac{d\omega}{d\Lambda} = & -\frac{8}{3d_c} \Bigl [ 
\Phi_2(f_\textsf{o},t,s,\omega) -\frac{1+s}{1+t}F_2(f_\textsf{o},t,s,\omega)
\notag \\
& +\frac{3\omega}{4}\Phi_0(f_\textsf{o},t,s,\omega) 
+ \frac{3\omega^2}{4(1+t)}F_0(f_\textsf{o},t,s,\omega)
\Bigr ] .
\label{eq:RG:omega}
\end{align}
We emphasize that $\omega$ is the ratio of the second and zeroth harmonics in the frame of reference where the second harmonic of $\varkappa$ is zero. 

Since the functions $F_2$ and $\Phi_2$ vanish at $\omega{=}0$, Eq. \eqref{eq:RG:omega} has the fixed point $\omega{=}0$. In the vicinity of this fixed point, 
$|\omega|{\ll} 1$ and for $f_\textsf{o}{\ll}1$, the RG equation for $\omega$ simplifies,
\begin{equation}
\frac{d\omega}{d\Lambda}=\frac{2}{ d_c}\mathcal{D}(t,s) f_\textsf{o}\,\omega  .
\label{eq:omega:RG:weak:f}
\end{equation}
We emphasize that the RG equations for $t,s,$ and $f_\textsf{o}$ remains intact to the linear order in $\omega$.

The function  $\mathcal{D}(t,s)$ has the following asymptotic expression in the limit of $|t|{\ll}1$ (see Appendix \ref{app: D(t,s)}):
\begin{equation}
\mathcal{D}(t,s) \simeq (1+s)\left(1 - \frac{119}{54}t\right) .
\label{eq:func:D}
\end{equation}
Solving Eq. \eqref{eq:omega:RG:weak:f} together with Eqs. \eqref{eq:RG:L1}, we find that $\omega$ flows towards the constant:
\begin{equation}
\omega\to \omega_\infty=\omega_0\Bigl [ 1+ f_\textsf{o}(0) \int\limits_0^{t_0} dt \frac{\mathcal{D}(t,s)}{g(t)}
e^{\int_{t_0}^t d\tau \overline{F}_0(\tau)/g(\tau)}\Bigr ].
\label{eq:omega:inf}
\end{equation}
Using asymptotic expansions \eqref{eq:asympt:g:F} and \eqref{eq:func:D}, we obtain from Eq. \eqref{eq:omega:inf} that 
$\omega_\infty{=}\omega_0[1{+}f_\textsf{o}(0)(1{+}s)]$ at $|t_0|{\ll}1$. 

We emphasize that since $f_\textsf{o}$ flows to zero a finite value of $\omega$ corresponds to zero second harmonic of disorder function, $\tilde{\psi}_2{=}0$. Therefore, the isotropic clean fixed point at $t{=}f_\textsf{0}{=}0$ is locally stable even if one starts away from invariant manifold.

At $f_\textsf{o}{\gg}1$ and $|\omega|{\ll}1$ Eq. \eqref{eq:RG:omega} can be simplified, 
\begin{equation}
\frac{d\omega}{d\Lambda}=\frac{2}{ d_c}\widetilde{\mathcal{D}}(t,s) \omega  .
\label{eq:omega:RG:strong:f}
\end{equation}
We note that the RG flow of the parameters $t$ and $s$ is unchanged within linear in $\omega$ approximation. The function $\widetilde{\mathcal{D}}(t,s)$ has the following asymptotic expression at $|t|, |s|{\ll}1$, 
\begin{equation}
\widetilde{\mathcal{D}}(t,s) \simeq \frac{20}{27}\left (t - \frac{19}{32}s\right ).
\end{equation}
Eq. \eqref{eq:omega:RG:strong:f} together with corresponding equations for $t$ and $s$ imply that $\omega$ flows towards a constant. At $f_\textsf{o}{\gg}1$ the RG flow for small magnitudes of $\omega$ is qualitatively the same as for $\omega{=}0$. 

We note that the parameter $\omega$ is related with the difference between $\Gamma$ and $\gamma$,
\begin{equation}
\omega=\frac{(1+s)(\Gamma^4-\gamma^4)}{3(\Gamma^4{+}\gamma^4)+2+3s(\Gamma^2-\gamma^2)^2}. 
\end{equation}
Also we mention that the renormalization group procedure described above implies a scale dependence of the parameters $\gamma$ and $\Gamma$. The flow of the orthorhombicity parameter $\gamma$ with the RG scale can be found from the following RG equation
\begin{equation}
\frac{d\gamma}{d\Lambda} = - \frac{4}{3d_c} \frac{1}{1+t}F_2(f_\textsf{o},t,s,\omega) \gamma .  
\label{eq:RG:gamma}
\end{equation}
We emphasize that the parameters $f_\textsf{o}, t, s,$ and $\omega$ in the right hand side of the above equation are governed by RG Eqs. \eqref{eq:RG:omega:full} and \eqref{eq:RG:omega}. The initial condition for Eq. \eqref{eq:RG:gamma} is $\gamma(0){=}(\varkappa_{xx}/\varkappa_{yy})^{1/4}$. As one can check, in accordance with Eq. \eqref{eq:RG:gamma}, the parameter $\gamma$ flows towards a constant.

In the case when Eq. \eqref{eq:relation:gamma} is not fulfilled the stretching factors are given by Eq. \eqref{eq:xiXY:full} with $\kappa$ depending on $\omega$. One can write down RG equations similar to Eqs. \eqref{eq:RG:stretching}. However, their analysis becomes too complicated. In the case of small deviations from the invariant manifold, when $|\omega|{\ll}1$ one can check that the temperature-driven and disorder-driven CT occurs qualitatively in the same way as for $\omega{=}0$. This means that the CT is almost insensitive to  
a weak breaking of the tetragonal crystal symmetry.

 \section{Discussions and conclusions\label{Sec:Disc}}

\subsection{Specific of RG flow in the presence of disorder}

In the presence of disorder the RG flow becomes many-parameterical one. It involves $t$, $s$, $f_\textsf{o}$, and $\gamma{-}\Gamma$ (parameter $\omega$). If the flow starts from the invariant manifold $\gamma{-}\Gamma{=}0$, it remains within it while $t$ and $f_\textsf{o}$ goes to zero whereas $s$ tends to the constant. The ultimate fate of the RG flow within the invariant manifold is the fixed point $t{=}f_\textsf{o}{=}0$. The properties of elastic response at this fixed point is similar to that of the clean anisotropic membrane. In particular, in the original frame of reference (before the affine transformation \eqref{eq:kxky:transform}) the bending rigidity and Young's modulus becomes \cite{Burmistrov2022}
\begin{equation}
\begin{split}
\varkappa(\bm{k}) & \sim (\gamma \cos^2\theta_{\bm{k}}+  \gamma^{-1} \sin^2\theta_{\bm{k}})^{2-\eta/2}(q_*/k)^\eta , \\
Y(\bm{k}) & \sim [(\gamma \cos^2\theta_{\bm{k}}+  \gamma^{-1} \sin^2\theta_{\bm{k}}) k^2/q_*^2]^{1-\eta} .
\end{split}  
\label{eq:bending:final:FP:1}
\end{equation}
Here we find the exponent $\eta{\simeq}2/d_c$ within the first order expansion in $1/d_c$. 

In spite of $f_\textsf{o}{=}0$ at the clean fixed point, the presence of disorder is reflected in different spatial behavior of two types of roughness correlation functions, (see Ref. \cite{Doussal2018} for a review)
\begin{equation}
\begin{split}
\overline{\langle \delta h(\bm{x}) \delta h(0)\rangle}\sim (\gamma x^2+ \gamma^{-1} y^2)^{1-\eta/2} ,\\
\overline{\langle h(\bm{x})\rangle \langle h(0)\rangle}\sim (\gamma x^2+ \gamma^{-1} y^2)^{1-\eta^\prime/2} .
\end{split}   
\label{eq:bending:final:FP:2}
\end{equation}
where $\eta^\prime{=}2\eta{\simeq}4/d_c$. The roughness correlation functions become dependent on the direction in $x$ - $y$ plane. We emphasize that a magnitude of the orthorhombicity parameter is determined by the initial values of the bending rigidity, $\gamma{=}(\varkappa_{xx}(0)/\varkappa_{yy}(0))^{1/4}$, and is not changed along RG flow within the invariant manifold. 

Now we discuss what happens if the RG flow starts away from the invariant manifold, i.e. at nonzero value of the difference $\gamma{-}\Gamma$. Our analysis of the RG flow in vicinity of the invariant manifold ($|\omega|{\ll}1$) suggests the following picture. The parameters $t$ and $f_\textsf{o}$ flow towards zero while $s$ tends to the constant. However, the RG flow leaves the plane $\gamma{-}\Gamma{=}\gamma(0){-}\Gamma(0)$ while the difference $\gamma{-}\Gamma$ flows towards a constant. Therefore, ultimately the RG flow ends at the clean fixed point $t{=}f_\textsf{o}{=}0$ and some values of $s$, $\gamma$, and $\Gamma$. Thus the bending rigidity, Young's modulus, and roughness correlation functions are given by Eqs. \eqref{eq:bending:final:FP:1}
and \eqref{eq:bending:final:FP:2}, respectively. But the parameter $\gamma$ is now determined by the solution of 
Eq. \eqref{eq:RG:gamma} and, thus, $\gamma$ depends on $t_0$, $s_0$, $\gamma(0)$, and $\Gamma(0)$. This implies that qualitatively, physical properties of 2D anisotropic flexible material, Eqs. \eqref{eq:bending:final:FP:1}
and \eqref{eq:bending:final:FP:2}, are independent of the fulfillment of the condition for the invariant manifold, $\gamma(0){=}\Gamma(0)$.

\subsection{Transition to the tubular crumpled phase}

Above we analize the CT for the case of the invariant manifold $\gamma(0){=}\Gamma(0)$.
In the clean case the transition to the tubular crumpled phase occurs at the temperature proportional to a typical value of the bending rigidity, cf. Eq. \eqref{eq:Tc:clean}. For $\varkappa_{xx}(0){>}\varkappa_{yy}(0)$, i.e. $\gamma{>}1$, the tubular crumpled phase corresponds to the vanishing stretching along $y$ axis, $\xi_y^2{=}0$. In the opposite case, $\gamma{<}1$, vice versa, the stretching along $x$ axis is zero, $\xi_x^2{=}0$. The presence of disorder reduces the transition temperature, cf. Eq. \eqref{eq:Tcr:f0:weak}. Interestingly, the initial slope of dependence of the transition temperature on a weak disorder is independent of the parameter $\gamma$, i.e. is the same for both tubular phases. 

At strong disorder $f_\textsf{o}{\gg}1$, the disorder-induced transition to the tubular crumpled phase occurs at critical disorder strength, cf. Eq. \eqref{eq:Bcr:st:small} and \eqref{eq:stretch:st:3}. For the line $s_0{=}t_0$ the critical disorder decreases monotonously with reduction of $t_0$. 
%
%
The critical disorder grows with increase of $T$ at low temperatures, cf. \eqref{eq:stretch:BT0}. The slope of the dependence of the critical disorder on $T$ is independent of the ortorhombicity parameter $\gamma$, i.e. is the same for both tubular phases. A positive slope (which is typical situation) suggests non-monotonous dependence of the critical disorder on temperature as in the isotropic case~\cite{Gornyi:2015a}.   

Away from the invariant manifold, $\gamma(0){\neq}\Gamma(0)$, the CT occurs qualitatively  in the same way as described above for the invariant manifold, $\omega{=}0$.

\subsection{Anomalous Hooke's law}

For a given stretching $\xi_\alpha$, a membrane tension can be computed as
\begin{equation}
\sigma_\alpha= \frac{1}{\xi_\alpha} \frac{\partial \mathcal{F}}{\partial \xi_\alpha} .
\label{eq:tension:def}
\end{equation}
Solving Eq. \eqref{eq:tension:def} for $\xi_\alpha$ at a given tension $\sigma_x$ applied along $x$ axis, we obtain the following anomalous Hooke's law at the clean isotropic fixed point, $t{=}f_\textsf{o}{=}0$ (see Sec. \ref{subsec: Isotrp}): 
\begin{equation}
    \delta \xi_x^2 \sim \gamma^{-1}(\sigma_x/\gamma)^{\alpha}, \quad  \delta \xi_y^2 \sim \gamma^{}(\sigma_x/\gamma)^{\alpha}.
    \label{eq: Hook}
\end{equation}
Here the exponent $\alpha {=} \eta/(2{-}\eta)$ (with $\eta{\simeq} 2/d_c$). 
We note that according Eq. \eqref{eq: Hook} the Hooke's law seems to have exactly the same form as in the clean case \cite{Burmistrov2022}. However, there is a subtlety. Eq. \eqref{eq: Hook} is indeed exactly the same if RG flow starts from the invariant manifold $\gamma{=}\Gamma$. If RG flow starts away from the invariant manifold the orthorhombicity parameter $\gamma$ should be found from solution of Eq. \eqref{eq:RG:gamma} upto the length scale induced by the tension, $L_\sigma{\sim}\sigma_x^{1/(2{-}\eta)}$. 
Neglecting subleading corrections at $\sigma{\to}0$ one can substitute $\gamma(L_\sigma)$ by $\gamma(\infty)$ in Eq. \eqref{eq: Hook}.

For strongly disordered membrane, $f_\textsf{o} {\gg} 1$, Hooke's law remains in the form of Eq. \eqref{eq: Hook}, but with a modified exponent $\alpha$:
\begin{equation}
    \alpha_{t,s \sim 0} = \tilde{\eta}/(2-\tilde{\eta}), \quad \alpha_{s \ne t \sim 1} = \hat{\eta}/(2-\hat{\eta}).
    \label{eq:alpha:strong:dis}
\end{equation}
where $\tilde{\eta} {\simeq} 1/(2d_c) $ and $\hat{\eta} {=}2 c_{F}/d_c$. We emphasize that the results \eqref{eq:alpha:strong:dis} describe transient regime while $L_\sigma$ is not too large (similar to the isotropic case  \cite{Gornyi2016}).

\subsection{Limitations of the $1/d_c$ expansion}\label{sec:dclim}

In this paper we consider lowest order of the $1/d_c$ expansion. Similar to the isotropic case, we find that the disorder parameter, $f_\textsf{o}$, flows always towards zero, i.e. the rippled (disorder-dominated) flat phase is marginal only. 

As known from the isotropic case \cite{Saykin2020b}, such an instability of the rippled flat phase is an artifact of the treatment with the lowest order in $1/d_c$. An account of the next order suggests the existence of the transition between flat rippled and flat clean phases at $f_\textsf{o}{\sim}d_c$. We expect that similar situation occurs in the anisotropic case. With that respect it would be interesting to extend our theory to the next order in $1/d_c$.

\subsection{The role of physical dimension, $D{=}2$} 

In this paper we focus on the case of physical dimension of a membrane, $D{=}2$. As was shown in Ref. \cite{Burmistrov2022} for the clean case, the fact that the orthorhombicity parameter $\gamma$ is not renormalized is specific for $D{=}2$. For $D{>}2$ there is a flow of $\gamma$ towards the isotropic case, $\gamma{\to}1$. This is related with the fact that the free energy after affine transformation depends on $\gamma$ even in the universal regime, $q{<}q_*$.  We expect that similar situation 
occurs in the disordered case for the invariant manifold. It's existence is limited to the physical dimension $D{=}2$ while for $D{=}2$ it does not exist. In analogy with the clean case, we expect that $\gamma$ and $\Gamma$ flow towards unity for $D{>}2$. It would we interesting to substantiate it by explicit calculations.

\subsection{Conclusions}

To summarize we developed the theory of anomalous elasticity in disordered 2D flexible materials with orthorhombic crystal symmetry. We demonstrated existence of infinitely many clean anisotropic flat phases. These phases have anisotropic bending rigidity and Young's modulus, cf. Eq. \eqref{eq:bending:final:FP:1}. However their scaling with the absolute value of momentum is the same as in a clean isotropic membrane. The disorder in these clean phases is responsible for  
different spatial behavior of various roughness correlation functions, cf. Eq. \eqref{eq:bending:final:FP:2}. In the clean flat phase these roughness correlation functions are anisotropic, cf. Eq. \eqref{eq:bending:final:FP:2} but with the same scaling with the distance as in the isotropic case. We found that the parameter $\gamma$ that distinguishes different flat phases may be not directly related with the degree of orthorhombicity of a membrane as it occurs in the clean case. 

With increase of disorder, $\mathcal{B}$, the clean flat phase undergoes the transition into the crumpled phase (see Fig.~\ref{pic: phasediagram}). The form of the transition curve $\mathcal{B}(T)$ resembles the corresponding curve for the crumpling transition in the isotropic disordered case. 
However, the crumpling transition occurs anisotropically so that a membrane crumples into a tubular phase. This disorder-driven transition 
is sensitive to the orthorhombicity parameter $\gamma$. The flat phase
corresponding to a particular magnitude of $\gamma$ undergoes transition to a tubular crumpled phase at a disorder strength which depends on $\gamma$, cf. Eq. \eqref{eq:Bcr:st:small}.
Our predictions are amenable to verification within numerical modeling of disorder-induced melting of flat phase in anisotropic atomic single layers.

\begin{acknowledgements}
We thank M. Glazov and V. Lebedev for useful comments and J. Schmalian for initial collaboration on the project.
The work was funded in part by the Russian Ministry of Science and Higher Educations, the Basic Research Program of HSE,  and by the Russian Foundation for Basic Research, grant No. 20-52-12019.
\end{acknowledgements}

\appendix

\begin{widetext}

\section{Asymptotic expressions for $F_{2m}$ and $\Phi_{2m}$ for $|t|$, $|s|\ll1$}\label{app: ast0}

Performing integrals in \eqref{eq: pifunc} for $\pi_{00}$, $\pi_{10}$, $\pi_{00}$, one can find: 
\begin{equation}
     \pi_{00}(\theta) \simeq \frac{1}{16 \pi}\left[1+\frac{2}{9}\cos(4\theta)t + \frac{85-2\cos(8\theta)}{90}t^2+ \frac{\cos(4\theta)(86+3\cos(8\theta))}{315}t^3\right]
 \end{equation}
 \begin{equation}
      \pi_{10}(\theta) \simeq \frac{1}{16 \pi}\left[1+\frac{1}{3}\cos(4\theta)t -\frac{1}{9}\cos(4\theta)s-  \frac{85-2\cos(8\theta)}{90}t s +\frac{85-2\cos(8\theta)}{45}t^2 \right]
 \end{equation}
 \begin{equation}
     \pi_{11}(\theta) \simeq \frac{1}{16 \pi}\left[1+\frac{4}{9}\cos(4\theta)t -\frac{2}{9}\cos(4\theta)s-\frac{80-2\cos(8\theta)}{45}t s  +\frac{125-3\cos(8\theta)}{45}t^2 -\frac{1}{18}s^2 \right]
 \end{equation}
One can see, there are no half harmonics $\cos(2\theta)$ in this expansions. After substitution into \eqref{eq:F2m:def}, \eqref{eq:Phi2m:def} and subsequent integration:
 \begin{multline}
    F_0 \approx \frac{1+3f_\textsf{o}+f_\textsf{o}^2}{(1+2f_\textsf{o})^2}+ \frac{(50f_\textsf{o}+269f_\textsf{o}^2+192f_\textsf{o}^3- 280f_\textsf{o}^4)}{81(1+2f_\textsf{o})^4}ts - \frac{(50+450f_\textsf{o}+1419f_\textsf{o}^2+1312f_\textsf{o}^3-240f_\textsf{o}^4)}{162(1+2f_\textsf{o})^4}t^2 +\\   \frac{(31 f_\textsf{o}^2+128f_\textsf{o}^3+120f_\textsf{o}^4)}{162(1+2f_\textsf{o})^4}s^2
\end{multline}
\begin{equation}
   F_4 \approx  -\frac{( 62 f_\textsf{o}^3+121f_\textsf{o}^2+66  f_\textsf{o}+11) t}{18 (2 f_\textsf{o}+1)^3}+\frac{ \left(40 f_\textsf{o}^3+44 f_\textsf{o}^2+11 f_\textsf{o}\right) s}{18 (2 f_\textsf{o}+1)^3}, \quad \Phi_4 \approx -\frac{(20 f_\textsf{o}^3+9 f_\textsf{o}^2) }{9 (2 f_\textsf{o}+1)^3}t +\frac{ (18 f_\textsf{o}^3 + 7f_\textsf{o}^2) }{18 (2 f_\textsf{o}+1)^3}s 
\end{equation}
 \begin{multline}
     \Phi_0 \approx \frac{f_\textsf{o}^2}{(2 f_\textsf{o}+1)^2}-\frac{(40  f_\textsf{o}^4 + 80  f_\textsf{o}^3+27  f_\textsf{o}^2 )}{162 (2 f_\textsf{o}+1)^4}s^2+ \frac{\left(480  f_\textsf{o}^4 +1000  f_\textsf{o}^3+383  f_\textsf{o}^2 \right) }{162 (2 f_\textsf{o}+1)^4} t^2- \frac{\left(320 f_\textsf{o}^4+560 f_\textsf{o}^3+203f_\textsf{o}^2\right)}{81 (2 f_\textsf{o}+1)^4}st
 \end{multline}
One can see, that $F_0, \; \Phi_0$ has only even degrees of anisotropy, and $F_4, \; \Phi_4$ has only odd degrees. 

\section{Asymptotic expressions for $\pi_{10}$ and $\pi_{11}$ for $1-|t| \ll 1$}\label{app: ast1}
For all further applications we need asymptotic expansions for only $\pi_{10}$ and $\pi_{11}$ in this case. For $F_{2m}$ and $\Phi_{2m}$ functions the main contribution in case $1-t \ll 1$ comes from $\theta \sim \pi/4$, $5\pi/4$ (see Eq. \eqref{eq: pifunc}), then we change the variables:  $\theta = \pi/4 + r \sqrt{1-t}$, $\varphi = \pi/4 + x \sqrt{1-t}$, $z = 1+ z(1-t)$, after making an expansion, $\pi_{10}$ takes the form: 
\begin{equation}
   \pi_{10} \simeq \frac{1}{12 \pi^2}\frac{1-\psi}{(1-\lambda)^{3/2}}\int\limits_{-\infty}^{+\infty}\frac{(x-r)^4}{(1+8x^2)^2}\frac{dxdz}{(x-r)^4+8(x-r)^2 z^2} = \frac{\sqrt{2}+4r \arctan(2\sqrt{2}r)}{192\pi}\frac{1-\psi}{(1-\lambda)^{3/2}} = \frac{1-\psi}{(1-\lambda)^{3/2}} h_1(r) 
\end{equation}
for $\pi_{11}$ in this way, we obtain:
\begin{equation}
   \pi_{11} \simeq \frac{1}{12 \pi^2}\frac{(1-\psi)^2}{(1-\lambda)^{5/2}}\int\limits_{-\infty}^{+\infty}\frac{(x-r)^8}{(1+8x^2)^2}\frac{dxdz}{((x-r)^4+8(x-r)^2 z^2)^2} =  \frac{(1-\psi)^2}{2(1-\lambda)^{5/2}} h_1(r) 
\end{equation}
After using equations \eqref{eq:F2mf:def}, \eqref{eq:Phi2mf:def} and making an expansion near $\theta_n \sim \pi/4 + \pi n/2 $, one can write \eqref{eq:F:Phi:0:0}. 

\section{Solution of RG eqs.  for $1-|t| \ll 1$}\label{app: solt1}
At first, we need to rewrite Eqs. \eqref{eq: RGtsim1} for variable $u =1-t$:
\begin{equation}\label{eq: RG :u}
    \frac{d u}{d\Lambda} \approx \frac{8}{3 d_c}c_{F}, \; u = u_0 + \frac{8}{3 d_c}c_{F}\Lambda, \quad \frac{d s}{d u} = -\frac{3}{4}\frac{c_{\Phi}}{c_{F}}\left(\frac{1}{3}+s\right) \frac{1-s}{u},\; \quad \frac{d f_\textsf{o}}{d u} \approx \frac{3}{4d_c} f_\textsf{o} \frac{c_{\Phi}}{c_{F}}\frac{1-s}{u}
\end{equation}
We can solve this equation analytically: 
\begin{equation}
    s = \frac{(u/u_0)^{-\alpha}(1+3s_0)/(1-s_0) -1}{3+  (u/u_0)^{-\alpha}(1+3s_0)/(1-s_0)}, \quad f_\textsf{o} = \frac{f_\textsf{o}(0)}{4} \left(1+ 3 s_0 + 3\left( \frac{u}{u_0}\right)^{\alpha}(1-s_0)\right)
\end{equation}
here one can see the logarithmic behaviour of the coupling constants.
\end{widetext}

\section{Transition at weak disorder} \label{weak-disorder}

 To solve Eq. \eqref{eq:RG:stretching} analytically we use a formal expansion of the RG Eqs. \eqref{eq:RG:2} to the first order in $f_\textsf{o}{\ll}1$. We shall use the following expansions
 \begin{equation}
 \begin{split}
    F_{2m}\simeq  \Bigl [ 1 - f_\textsf{o} & - f_\textsf{o} (s-t)\partial_t\Bigr ]  \overline{F}_{2m} + O(f_\textsf{o}^2) , \\
   & \Phi_{2m}\simeq O(f_\textsf{o}^2) ,
   \end{split}
 \end{equation}
 Then, we find (cf. Eqs. \eqref{eq:RG:L1})
 \begin{align}
\frac{dt}{d\Lambda}  = & -  \frac{2}{d_c}
\Bigl [g(t)(1-f_\textsf{o})+(s-t)(\overline{F}_0(t)-g^\prime(t))f_\textsf{o} \Bigr ] ,\notag \\
\frac{df_\textsf{o}}{d\Lambda}  = &
-  \frac{2}{d_c}\overline{F}_0(t) f_\textsf{o}, \qquad \frac{ds}{d\Lambda} = 0,
\label{eq:RG:f0:small}\\
\frac{d\ln \varkappa_0}{d\Lambda}  = & \frac{2}{d_c} \Bigl [\overline{F}_0(t) (1-f_\textsf{o})-(s-t)\overline{F}_0^\prime(t)f_\textsf{o}\Bigr ] . \notag 
\end{align}
Solving Eqs. \eqref{eq:RG:stretching} together with Eqs. \eqref{eq:RG:f0:small}, we obtain the transition temperature  to the tubular phase with $\xi_y^2=0$ ($\alpha {=}X,Y$)
\begin{equation}
T_{\rm cr,\alpha} = T_{\rm cr,\alpha}^{(0)}
 - \mathcal{A}(t_0,s_0) \mathcal{B}(0) \sqrt{\varkappa_{xx}(0)\varkappa_{yy}(0)} .
\label{eq:Tcr:f0:weak}
\end{equation}
Here  $\varkappa_0(0)$, $f_\textsf{o}(0)$, and $t_0$
denote initial values of the corresponding variables at the ultra-violet momentum scale given by the inverse Ginzburg length $q_*$. 
$T_{\rm cr,\alpha}^{(0)}$ denotes the transition temperature in the absence of disorder. For $\gamma{>}1$ it is given as  \cite{Burmistrov2022}
\begin{equation}
T_{\rm cr,Y}^{(0)} = \frac{8\pi}{d_c^2} 
\bigl (\varkappa_{xx}(0)\varkappa_{yy}^3(0)\bigr )^{1/4}\, \overline{f}(t_0)  .
\label{eq:Tc:clean}.
\end{equation}
The coefficient $\mathcal{A}(t_0,s_0)$ determines the slope of the dependence of the critical temperature on $\mathcal{B}(0)$,
\begin{gather}
\mathcal{A}(t_0,s_0) {=} \frac{1+t_0}{1+s_0} \int\limits_{0}^{t_0} \frac{dt\, e^{-\int_{t}^{t_0} d\tau \frac{ \overline{F}_0(\tau)}{g(\tau)}}}{g(t)\sqrt{1-t^2}} 
\Biggl\{
e^{-\int_{t}^{t_0} d\tau \frac{ \overline{F}_0(\tau)}{g(\tau)}}\notag \\
\times
\Bigl[ 1 +\frac{1- s_0 t}{1-t^2}
+ (s_0-t) \frac{g^\prime(t)-\overline{F}_0(t)}{g(t)}  \Bigr ]
 +  \int \limits_{t}^{t_0} d\tau
(s_0-\tau) 
\notag \\ \times
e^{-\int_{\tau}^{t_0} d\tau_1 \frac{ \overline{F}_0(\tau_1)}{g(\tau_1)}}
\Biggl [ \left (\frac{\overline{F}_0(\tau)}{g(\tau)} \right )^\prime + \left (\frac{\overline{F}_0(\tau)}{g(\tau)} \right )^2 \Biggr ]
\Biggr \} \notag \\ \Biggl / \Biggl [\int\limits_{0}^{t_0} \frac{dt\, e^{-\int_{t}^{t_0} d\tau \frac{ \overline{F}_0(\tau)}{g(\tau)}}}{g(t)\sqrt{1-t^2}} \Biggr ] .
\label{eq:def:mathA}
\end{gather} 
We emphasize that the coefficient $\mathcal{A}(t_0,s_0)$ is independent of $1/d_c$. Thus correction to the critical temperature due to disorder has no smallness in $1/d_c$. Also it is independent of the parameter $\gamma$, i.e. it is symmetric with respect to interchange of $x$ and $y$.

General expression \eqref{eq:def:mathA} can be simplified for $|t_0|{\ll}1$:
 \begin{gather}
     \mathcal{A}(t_0,s_0)= \frac{1{+}t_0{-}54
     s_0 t_0/119}{1+s_0} . 
 \end{gather}
 As one can see, disorder reduces the transition temperature for $|t_0|{\ll}1$. 
 
\begin{figure}
\centering
		  	\includegraphics[width=0.8\columnwidth]{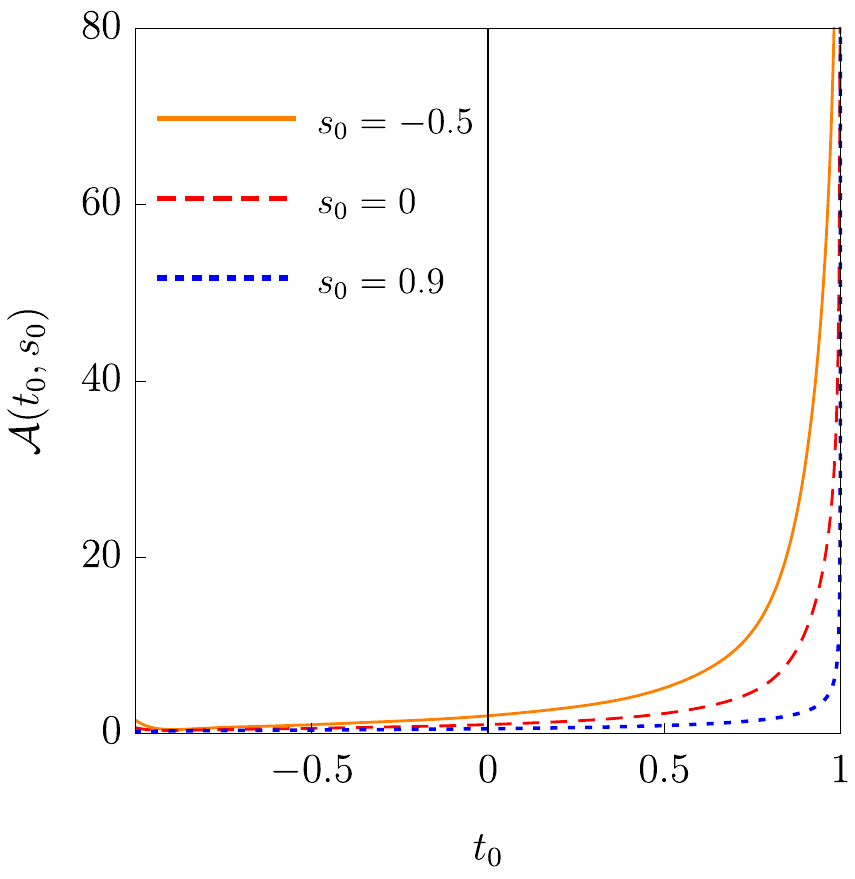}
		  	\caption{The function $\mathcal{A}(t_0,s_0)$, which determines the slope of the critical curve $T(\mathcal{B})$ near $T_c$}\label{pic: Afunc} 
\end{figure}

 In the case of strong anisotropy, $1{-}t_0 {\ll} 1$ the expression \eqref{eq:def:mathA} can be written as
   \begin{equation}
     \mathcal{A}(t_0,s_0) \simeq  
     \frac{1-s_0}{1+s_0} \frac{c_{\mathcal{A}}}{\sqrt{1-t_0}} ,
 \end{equation}
 where 
 $c_{\mathcal{A}}{\approx} 1.38$ we found from numerical calculation. We assume that $s_0$ is not too close to the unity, $1{-}s_0{\gg} 1{-}t_0$. As one can see, 
for $t_0$ close to the unity, 
 the critical temperature \eqref{eq:Tcr:f0:weak} becomes particular sensitive to disorder.  
Numerical analysis of Eq. \eqref{eq:def:mathA} suggests that $\mathcal{A}(t_0,s_0){>}0$ for $|t_0|, |s_0| {<}1$ (see Fig.~\ref{pic: Afunc}), i.e. weak disorder always decreases transition temperature. 

\section{Temperature dependence of the critical disorder\label{T-correction}}

The critical disorder 
slightly depends on temperature. Specifically, at not too high  temperatures, it   acquires linear in $T$ correction ($\alpha{=}X,Y$),
\begin{equation}
\mathcal{B}_{\rm cr, \alpha}(T) \simeq   \mathcal{B}_{\rm cr, \alpha} \Bigl [1 + \mathcal{C}(t_0,s_0)  T/T_{\rm cr, \alpha} \Bigr ].
\label{eq:stretch:BT0}
\end{equation}
The linear-in-$T$ correction appears due to both the direct contribution (proportional to $T$) in Eqs. \eqref{eq:RG:stretching} for stretching factors and the $1/f_\textsf{o}$ corrections to the RG equations at $f_\textsf{o}{\gg} 1$. Typically, the later is larger than the former such that $\mathcal{C}(t_0,s_0){>}0$. The analytical expression for the slope of the critical disorder with temperature, $\mathcal{C}(t_0,s_0)$, is too cumbersome. Instead, it is more convenient to extract it directly from numerical solution of the RG Eqs. \eqref{eq:RG:2} and Eqs. \eqref{eq:RG:stretching}. 

To illustrate Eq. \eqref{eq:stretch:BT0}, we consider vicinity of the fixed point at $t{=}s{=}0$. To the linear order in $s$ and $t$, Eqs. \eqref{eq:RG:stretching} becomes the same as in isotropic case
\begin{equation}
\begin{split}
     \frac{d\xi^2_x}{d\Lambda} = - \frac{d_c}{4 \pi \gamma} \frac{T (1+f_\textsf{o})}{\varkappa_0} ,\\
     \frac{d\xi^2_y}{d\Lambda} = - \frac{d_c \gamma}{4 \pi}\frac{T (1+f_\textsf{o})}{\varkappa_0} .
     \end{split}
\end{equation}
Using the symmetry relations \eqref{eq:symm:F:Phi}, we obtain to the linear order in $s$ and $t$,
\begin{equation}
\frac{d \varkappa_0}{d \Lambda} = \frac{\varkappa_0}{2d_c} \left(1+ \frac{2 }{f_\textsf{o}}\right), \quad \frac{d f_\textsf{o}}{d \Lambda} = -\frac{3}{2d_c} .
\end{equation}
Solving the above equations, one finds the critical disorder in the form of Eq. \eqref{eq:stretch:BT0} with the slope 
\begin{equation}
\mathcal{C}(t_0,s_0)\simeq 16, \quad |t_0|\ll 1,\, |s_0|\ll 1.     
\end{equation}

\color{black}

\begin{widetext}
\section{Derivation of $\mathcal{D}(t,s)$}\label{app: D(t,s)}
One can find behaviour of functions $F_{2m}$, $\Phi_{2m}$ away from invariant manifold in limit of $f_\textsf{o} \ll 1$, $|\omega| \ll 1$:
\begin{equation}
    F_{2m} \simeq \frac{1}{16 \pi} \int\limits_{0}^{2 \pi}\frac{d \theta}{2 \pi} \frac{\cos(2m \theta)}{\kappa(t,\theta) \pi_{00}(\theta)} + f_\textsf{o} \frac{1}{16 \pi} \int\limits_{0}^{2 \pi}\frac{d \theta}{2 \pi} \frac{\cos(2m \theta)}{\kappa(t,\theta) \pi_{00}(\theta)}\left(\frac{1+\omega \cos(2\theta)+ s\cos(4\theta)}{\kappa(t,\theta) }- 2 \pi_{10}(\theta)\right), \; \Phi_{2m}\simeq 0 
\end{equation}
After substitutions in \eqref{eq:RG:omega} we obtain:
\begin{equation}
    \frac{d\omega}{d\Lambda} \simeq  \frac{8}{3d_c} 
\frac{1+s}{1+t}F_2(f_\textsf{o},t,s,\omega) = \frac{2}{d_c} \mathcal{D}(t,s) f_\textsf{o} \omega 
\end{equation}
where $\mathcal{D}(t,s)$ defined as:
\begin{equation}
    \mathcal{D}(t,s) = \frac{4}{3} \frac{1+s}{1+t} \int\limits_{0}^{2 \pi}\frac{d \theta}{32 \pi^2} \frac{\cos(2\theta)}{(1+t \cos(4\theta)) \pi_{00}(\theta)}\Bigl [\frac{\cos(2 \theta)}{1+ t\cos(4\theta)}- 2 \tilde{\pi}_{10}\Bigr]
\end{equation}
where $\tilde{\pi}_{10}$ has only the second harmonics contribution.
\end{widetext}

\bibliography{biblio-disorder-anisotropy}

\end{document}